\documentclass[aps,prd,11pt,notitlepage,nofootinbib]{revtex4}
\usepackage{epsfig}
\newcounter{fig}   \newcommand{\lbfig}[1]{\refstepcounter{fig}
\label{#1} }
\usepackage{bm}

\usepackage{amsmath}
\usepackage{amsfonts}
\usepackage{graphicx}
\usepackage{amssymb}
\usepackage{amssymb}
\usepackage{latexsym}
\usepackage{hyperref}
\usepackage{array}
\usepackage{bbold}

\newcommand{\bea}{\begin{eqnarray}}
\newcommand{\eea}{\end{eqnarray}}
\newcommand{\be}{\begin{equation}}
\newcommand{\ee}{\end{equation}}
\newcommand{\re}[1]{(\ref{#1})}

\usepackage{xcolor}

\begin{document}
\title{Fermions on the kink revisited}
\begin{abstract}

We study fermion modes localized on the kink in the 1+1 dimensional $\phi^4$ model,
coupled to the Dirac fermions with backreaction. Using numerical methods we construct self-consistent
solutions of the corresponding system of coupled integral-differential equations
and study dependencies of the scalar field of the kink and the
normalizable fermion bound states on the values of the values of the parameters of the model. We show that
the backreaction of the localized fermions significantly modifies the solutions, in particular
it results in spatial oscillations of the profile of the kink and violations
of the reflection symmetry of the configuration.
\end{abstract}
\author{Vladislav Klimashonok}
\author{Ilya Perapechka}
\affiliation{Department of Theoretical Physics and Astrophysics,BSU, Minsk 220004, Belarus}
\author{Yakov Shnir}
\affiliation{BLTP, JINR,\\Dubna 141980, Moscow Region, Russia,\\
}

\maketitle
\section{Introduction}
Many non-linear physical system support solitons,
spatially localized field configurations with various ramifications for
condensed matter physics, non-linear optics, nuclear physics, cosmology and
quantum field theory \cite{Solitons,Dauxois,Manton:2004tk,Shnir2018,Vilenkin}.
Perhaps the simplest examples of solitons are the kinks which appear in
models in one spatial dimension with a potential possesing two or more
degenerated minima, see, e.g., \cite{Vachaspati,Panos}. Double well potential corresponds
to the nonintegrable $\phi^4$ model, another interesting example are
the kinks in the $\phi^6$ theory \cite{Lohe:1979mh}, kink-antikink collision in this model
have attracted much attention recently \cite{Dorey:2011yw,Gani:2014gxa}.

A peculiar feature of topological solitons is the relation between the topological charge of the
configuration and the number of fermion zero modes localized on the soliton.
The Atiyah-Patodi-Singer index theorem \cite{APS} yields a remarkable relation between these quantities.

Fermionic zero modes of the solitons were discussed first in the pioneering work \cite{zeromode},
similar localized states exist on the vortices \cite{Jackiw:1981ee}, domain walls \cite{Stojkovic:2000ub},
monopoles \cite{Rubakov:1982fp,Callan:1982au},
sphalerons \cite{Nohl:1975jg,Boguta:1985ut} and
skyrmions \cite{Hiller:1986ry,Kahana:1984dx,Kahana:1984be,Ripka:1985am}.
The presence of localized fermion modes leads to some very interesting phenomena such as
monopole catalysis of proton decay \cite{Rubakov:1982fp,Callan:1982au},
possible existence of superconducting cosmic strings \cite{Witten:1984eb} and appearance of
fractional quantum numbers of solitons \cite{Jackiw:1975fn,Jackiw:1981ee}.

Fermions bounded by kinks were considered in many papers
\cite{Dashen:1974cj,Jackiw:1975fn,Chu:2007xh,Liu:2008pi,Brihaye:2008am}.
However, nearly all of the studies neglected the backreaction of the fermions on the soliton, moreover
only zero modes were considered in most cases. There has however been some attempt
to take into account the back-reaction of the fermion on the kink \cite{Gani:2010pv,Mohammadi:2014kca},
although self-consistent solution is still missing. One of the main reasons for that is the
enormous computational complexity of the problem, there is no analytical solution of the corresponding system of coupled
integral-differential equations.

A main objective of this paper is to reconsider this system consistently.
Recently,  we developed new numerical scheme which was successfully applied to
examine the effects of backreaction of localized fermionic modes on planar skyrmions
\cite{Perapechka:2018yux,Perapechka:2019dvc}.
We have found that there is a tower of fermionic modes of two different types,
localized by the soliton with one level crossing
mode. Furthermore, in \cite{Perapechka:2019dvc} we discussed a novel mechanism of exchange interaction between
the skyrmions and constructed stable multi-soliton configurations bounded by the attractive interaction
mediated by the chargeless fermionic modes.

In the present paper we revisit the fermion-kink bounded system with backreaction.
Apart the well known zero mode, which does not affect the kink for any values of the Yukawa coupling,
we find various localized fermion modes with finite energy. The number of these bound modes
increases as the Yukawa coupling becomes stronger, they are linked to the states of positive and negative continuum.
We find that, as we increase the coupling, the effects of backreaction of the fermions on the kink becomes more and more
significant. Furthermore, the localized fermions may give rise to additional exchange interaction between the solitons.

This paper is organised as follows. In Section II we present the $\phi^4$ model coupled to Dirac fermions via the usual Yukawa
coupling. Numerical results are presented in Section III, where we
describe the solutions of the model and discuss the spectral flow of the localized fermionic states with backreaction on the
kink. Conclusions and remarks are formulated in the last Section.

\section{The model}
We consider a coupled fermion-scalar system in 1+1 dimensions
defined by the Lagrangian
\begin{equation}
\mathcal{L}=\frac{1}{2}\partial_\mu\phi\partial^\mu \phi
+ \bar \psi\left[ i\gamma^\mu \partial_\mu  -m - g\phi \right]\psi
-U(\phi) \, ,
\label{lag}
\end{equation}
where $U(\phi)$ is a potential of the self-interacting scalar field,
$\psi$ is a two-component spinor and $m,g$ are the bare
mass of the fermions and the dimensionful Yukawa coupling constant,
respectively. The matrices $\gamma_\mu$ are
$\gamma_0=\sigma_1$, $\gamma_1=i\sigma_3$ where $\sigma_i$ are the Pauli matrices, and
$\bar \psi = \psi^\dagger\gamma^0$.
The $\phi^4$ model corresponds to the quartic potential $U(\phi)=\frac12 \left(1-\phi^2\right)^2$
with two vacua $\phi_0\in \left\{-1,1 \right\}$.

The field equations of the system are given by
\be
\begin{split}
&\partial_\mu\partial^\mu \phi +g\bar \psi \psi - 2 \phi +2 \phi^3 = 0\,;\\
& i\gamma^\mu \partial_\mu \psi  - m\psi  - g\phi \psi = 0\, .
\end{split}
\label{eqs}
\ee
Using the usual parametrization of
a two-component spinor
$$
\label{ansFer}
\psi= e^{-i\epsilon t}\left(
\begin{array}{c}
u(x)\\
v(x)
\end{array}
\right)\, ,
$$
we obtain the following coupled system of static equations
\be
\begin{split}
\phi_{xx} + 2g uv -2\phi+2\phi^3 &=0\, ;\\
u_x+(m+g\phi)u&=\epsilon v\, ;\\
-v_x+(m+g\phi)v&=\epsilon u\, .
\end{split}
\label{eq2}
\ee
This system is supplemented by the normalization condition $\int\limits_{-\infty}^\infty dx~(u^2+v^2) =1$, thus
the configuration as a whole could can be characterized by two quantities, the fermionic density distribution
$\rho_f=u^2+v^2$ and the topological density, i.e. the profile of the scalar field of the kink $\phi(x)$.

Note that the first equation in the system of dynamical equations  \re{eq2} enjoys the reflection symmetries
\be
x\to -x,\;\; \phi\to-\phi, \;\;uv\to -uv,
\label{symmetry1}
\ee
while the equations on the spinor components coupled to the scalar field, are invariant with respect to the
transformations
\be
x\to -x,\;\;u\to v,\;\; v\to u
\label{symmetry2}
\ee

Consideration of the fermionic modes is usually related with
simplifying assumption that the scalar field background is fixed
\cite{Dashen:1974cj,Jackiw:1975fn,Chu:2007xh,Liu:2008pi,Brihaye:2008am}.
In the decoupled limit $g=0$,
the $\phi^4$ model supports a spatially localized static topological soliton, the kink:
\begin{equation}
 \phi_K(x)=\tanh (x-x_0)
 \label{kink}
\end{equation}
interpolating between the vacua $\phi_0=-1$ and $\phi_0=1$.
Here $x_0$ is the position of the center of kink. The antikink solutions can be found by
the inversion $x\to -x$. Clearly, the kink field is parity-odd, it agrees with the
symmetry condition \re{symmetry1}. Then the reflection symmetry of the Dirac equation \re{symmetry2}
means that the positive energy fermionic states localized on the kink are also the
negative energy states localized on the antikink, and vice versa \cite{Chu:2007xh}.
Further, due to this symmetry
there is only one zero mode of the
Dirac equation, which does not depend on the value of the Yukawa coupling
$g$ \cite{Dashen:1974cj,Jackiw:1975fn,Chu:2007xh}
\be
\psi_0=N_0\left(
\begin{array}{c}
\frac{e^{-mx}}{\cosh^g x}\\
0
\end{array}
\right)\, ,
\label{exact-zero}
\ee
where $N_0$ is a normalization factor. In the special case of the $N=1$
supersymmetric generalization of the model \re{lag} \cite{DiVecchia:1977nxl}
this mode is generated via the supersymmetry transformation of the boson field of the static kink.

It was noticed that further increase of the Yukawa coupling $g$  gives rise to other fermion modes
with non-zero energy, which are localized on the kink \cite{Dashen:1974cj,Chu:2007xh}.
Indeed, the system of two first order differential equations in \re{eq2}
can be transformed into two decoupled
second order equations for the components $u$ and $v$ \cite{Dashen:1974cj}
\be
\begin{split}
-u_{xx}+\left( (m+g \phi)^2 - g \phi_x \right) u&=\epsilon^2 u\, ;\\
-v_{xx}+\left( (m+g \phi)^2 + g \phi_x \right) v&=\epsilon^2 v\, .
\end{split}
\label{eq3}
\ee
They are Schr\"odinger-type equations, for
the fermions in the external static background field of the kink \re{kink}
the corresponding potential is
\be
U_f=(m+g \tanh x)^2 \pm \frac{g}{\cosh^2 x}
\label{pot-fermi}
\ee
In the limit of zero bare mass of the fermions, $m=0$, the potential \re{pot-fermi}
becomes reduced to the usual P\"oschl-Teller potential, so
the equations \re{eq3} can be solved analytically \cite{Dashen:1974cj,Chu:2007xh}. Further,
it was pointed out that as the Yukawa coupling $g$ increases, the potential well
becomes deeper and new levels appear in the
spectrum of the bound states \footnote{Interestingly, the corresponding equations for the fermions
on the background of static kinks of the completely integrable sine-Gordon model, or for the fermions on the
kinks of the $\phi^6$ model with triple vacuum \cite{Lohe:1979mh,
Dorey:2011yw}, do not support localized states with non-zero eigenvalues,
there are only zero modes in the spectrum.}. For example, there is a bound state solution for the
massless fermions, which appears as the coupling increases above $g_{cr}=1$,
\be
\psi_1=N_1\left(
\begin{array}{c}
\frac{\pm \sqrt{2g-1}\tanh x}{\cosh^{g-1} x}\\
\frac{1}{\cosh^{g-1} x}
\end{array}
\right)\, ,
\label{exact-first}
\ee
with eigenvalues $\epsilon_1= \pm \sqrt{2g-1}$ \cite{Postma:2007bf,Chu:2007xh,Gani:2010pv}.
Other solutions also can be written in a closed form, see \cite{Chu:2007xh}.

However, as the coupling becomes stronger, the backreaction of the bounded
modes  could significantly affect the scalar field, so the analytical solution
for the fermion modes bounded by the kink in not self-consistent for large values of $g$.
Indeed, as we will see below, at strong coupling the
exact self-consistent numerical solutions of the coupled system of equations \re{eqs}
become very different from the analytical results for the fermions in the external field of the static kink.
Our goal here is to investigate this effect in a systematic way.

\section{Numerical results}
We have solved numerically the full system of integral-differential equation \re{eqs} with
the normalization condition on the spinor field using 8th order finite-difference method.
The system of equations is discretized on a uniform grid with usual size of 5000 points.
To simplify our calculations, we consider only positive semi-infinite line taking into account the symmetry of the
configuration \re{symmetry1},\re{symmetry2}.
Further, we map semi-infinite region onto the unit interval $\left[0,1\right]$
via the coordinate transformation $\tilde{x}=\frac{x}{c + x}$. Here $c$ is an arbitrary constant which is used to
adjust the contraction of the grid.
The emerging system of nonlinear algebraic equations is solved
using a modified Newton method.
The underlying linear system is solved with the Intel MKL PARDISO sparse direct solver.
The errors are on the order of $10^{-9}$.

To obtain numerical solutions of the system \re{eqs} we have to impose appropriate
boundary condition for the spinor field, both at the center of the kink and at the vacua.
For fermions localized on the kink, we have to impose
$$
\phi\big|_{-\infty}=-1,\;\quad \phi\big|_\infty=1, \; \quad
u\big|_{-\infty}=u\big|_{\infty}=v\big|_{-\infty}=v\big|_{\infty}=0 \, .
$$

First, we consider the normalized fermions with zero bare mass $m=0$.
Taking into account the symmetry properties \re{symmetry2} and the
linearized equations \re{eq2} for the spinor field at $x=x_0=0$, we can
can classify the corresponding solutions according to their parity.
Thus,  we consider two types of the
boundary conditions for the massless fermions at the center of the kink
$$
u_x \big|_{x_0}=0\, \qquad v\big|_{x_0}=0 \quad {\rm or}\quad u\big|_{x_0}=0\, \qquad v_x\big|_{x_0}=0 \, .
$$
We will refer to the modes of the first type to as $A_k$-modes and to the modes of the second tape to as
$B_k$-modes, i.e. the modes of the type $A$ have symmetric $u$-component and anti-symmetric $v$-component,
while the modes of the type $B$ have anti-symmetric $u$-component and symmetric $v$-component.
Here the index $k$ corresponds to the minimal number of nodes of the components, for example the zero
mode \re{exact-zero} is denoted as $A_0$. Note that for all modes, the number of nodes of the component $u$ is one
node more than the number of nodes of the component $v$.

\begin{figure}[t]
    \begin{center}
        \includegraphics[width=.435\textwidth, trim = 40 20 90 20, clip = true]{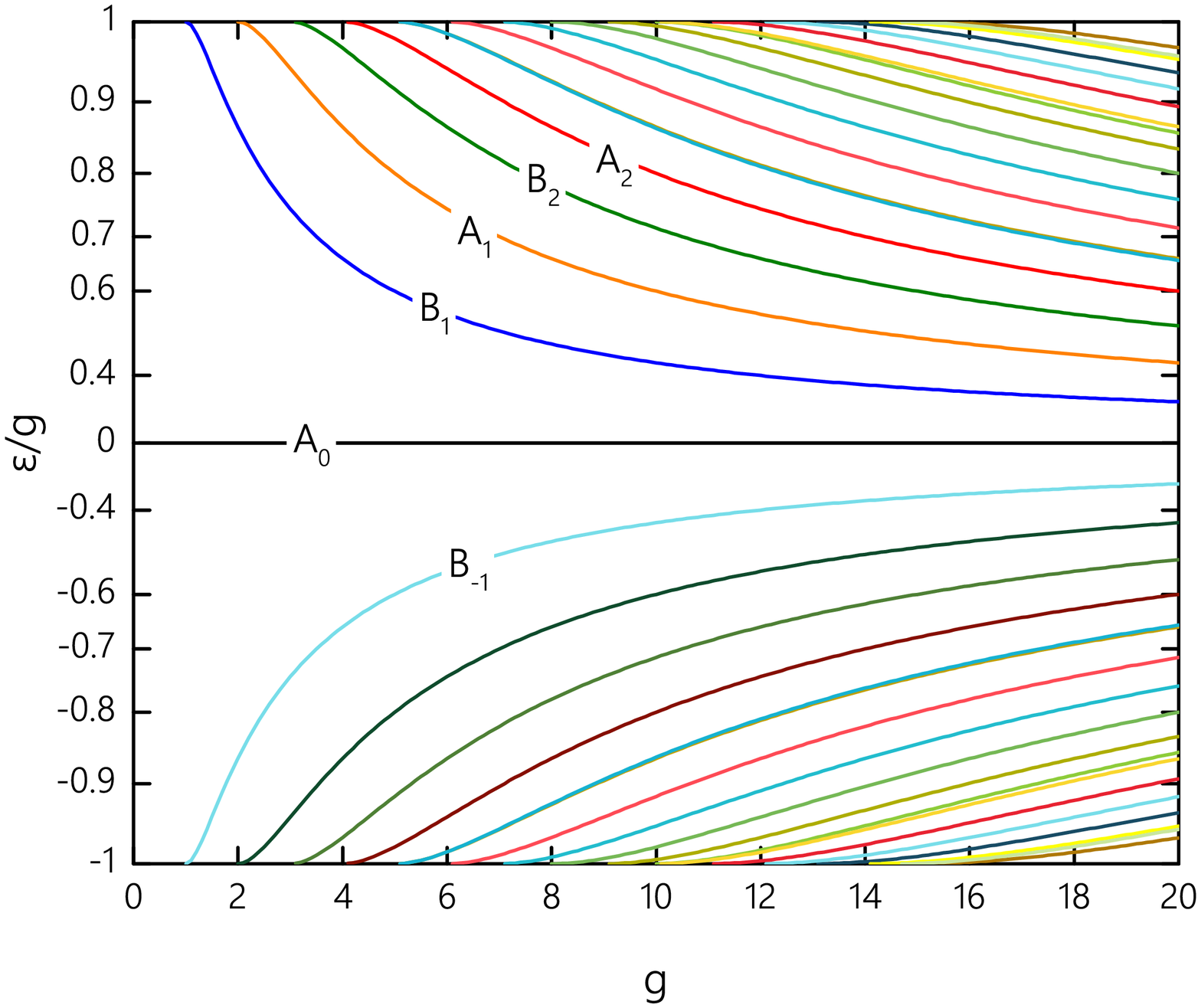}
        \includegraphics[width=.435\textwidth, trim = 40 20 90 20, clip = true]{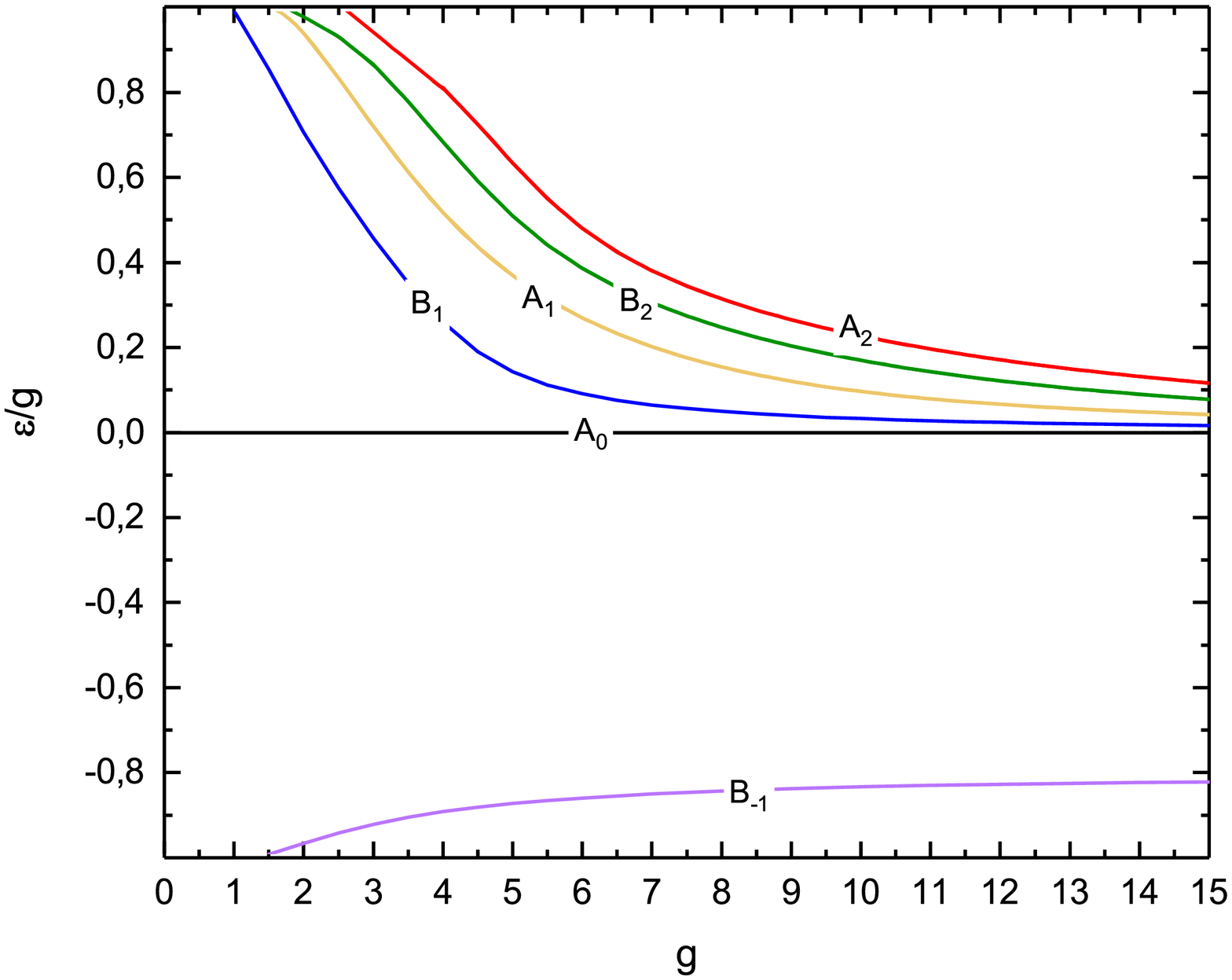}
    \end{center}
    \caption{\small
Normalized energy  $\frac{\epsilon}{g}$ of the localized fermionic states
as a function of the Yukawa coupling $g$
for several fermion modes at $m=0$ without backreaction (left) and with backreaction of the fermions
on the kink (right).
}
\lbfig{Fig1}
\end{figure}

In the decoupling limit the backreaction of the fermions on the kink is neglected, the pair of the first order
equations on the spinor components in the system \re{eqs}
describes the fermion states in the external
scalar field of kink $\phi_K(x)$ \re{kink}. In such a case the energy spectrum of the localized fermions
is symmetric with respect to inversion $\epsilon \to -\epsilon$,
apart zero mode $A_0$, each state with a positive eigenvalue $\epsilon$ has a
counterpart with reflected
anti-symmetric $u(v)$-component and a negative eigenvalue $-\epsilon$, see Fig.~\ref{Fig1}, left plot.

\begin{figure}[t!]
    \begin{center}
        \includegraphics[width=.435\textwidth, trim = 40 20 90 20, clip = true]{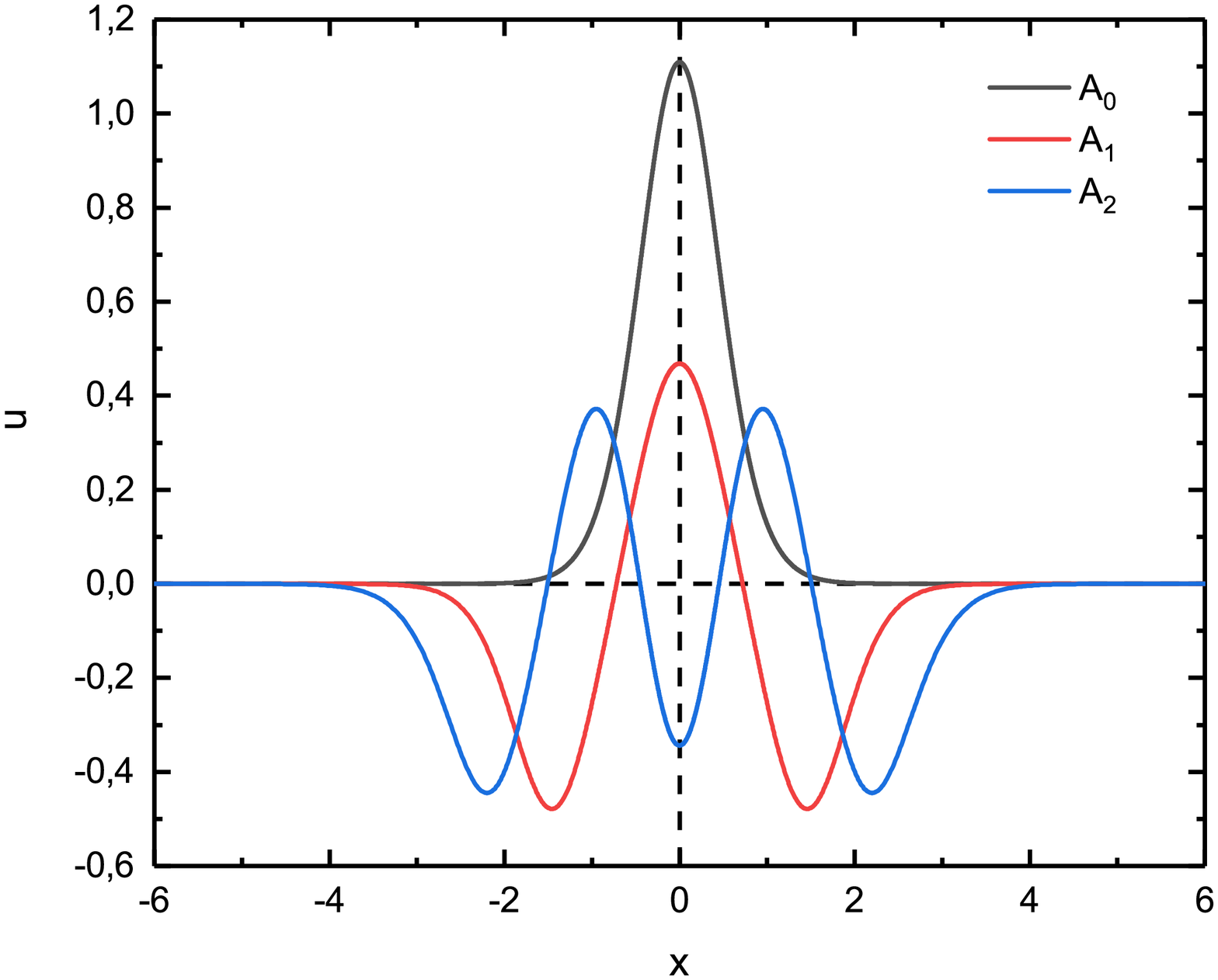}
        \includegraphics[width=.435\textwidth, trim = 40 20 90 20, clip = true]{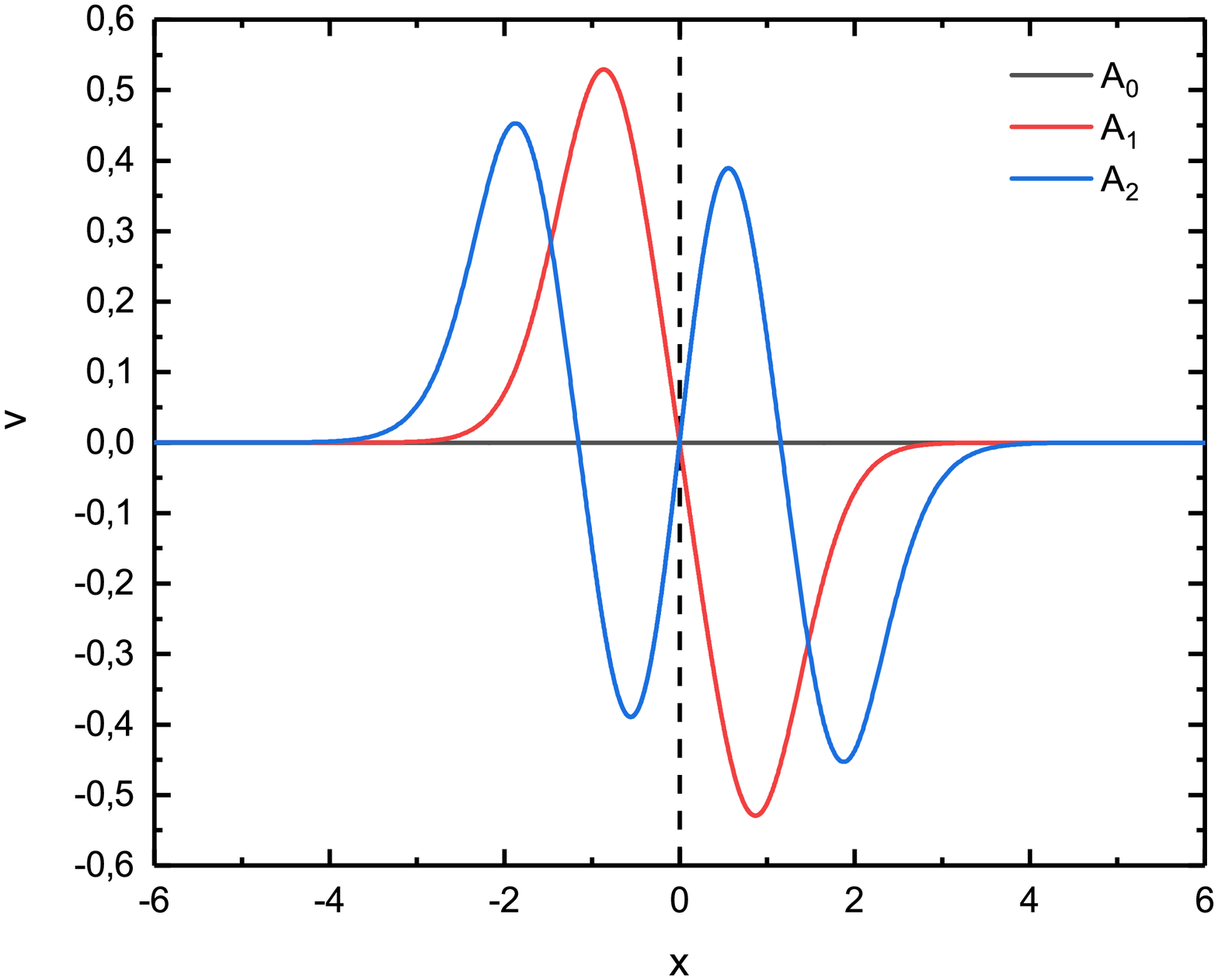}
        \includegraphics[width=.435\textwidth, trim = 40 20 90 20, clip = true]{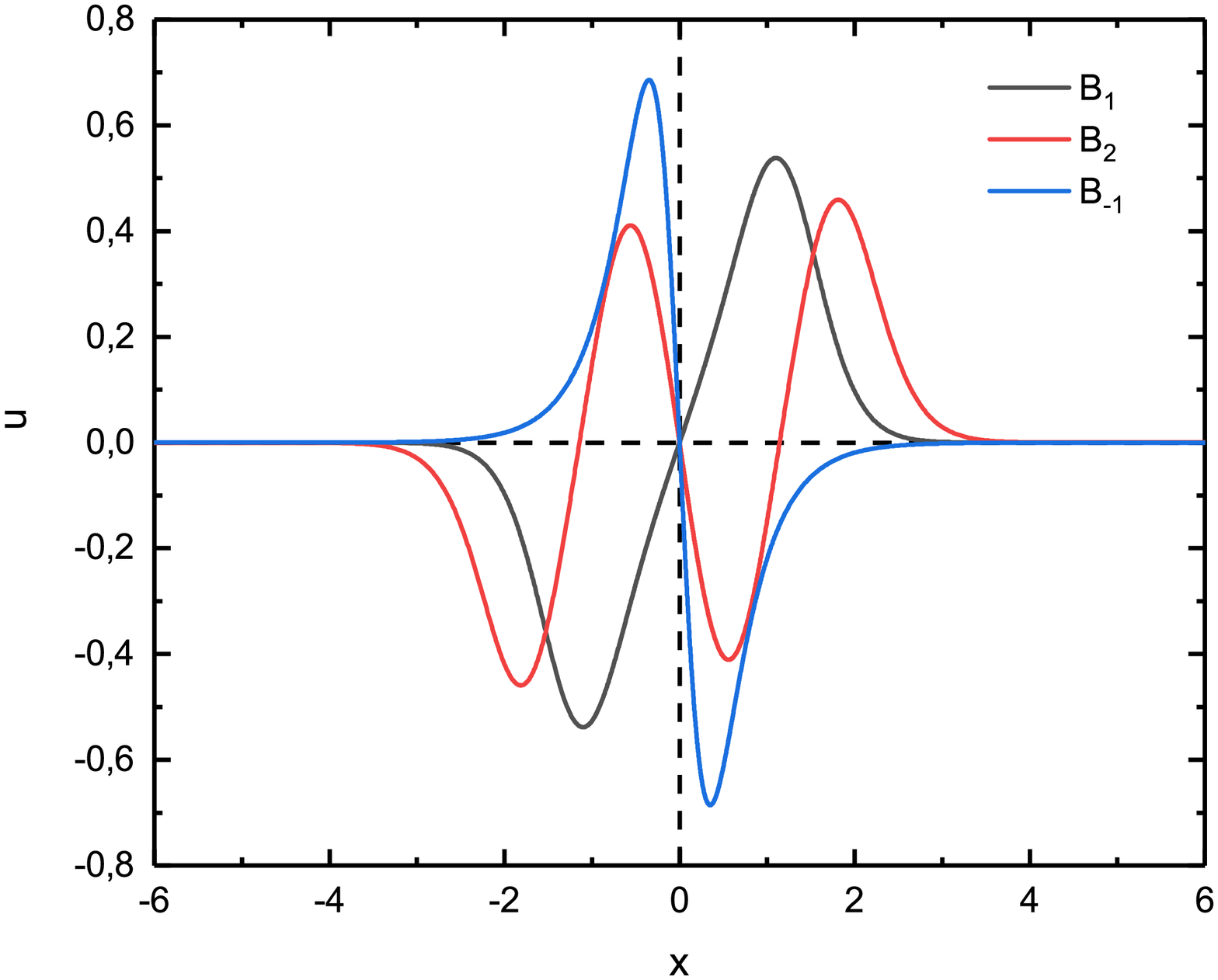}
        \includegraphics[width=.435\textwidth, trim = 40 20 90 20, clip = true]{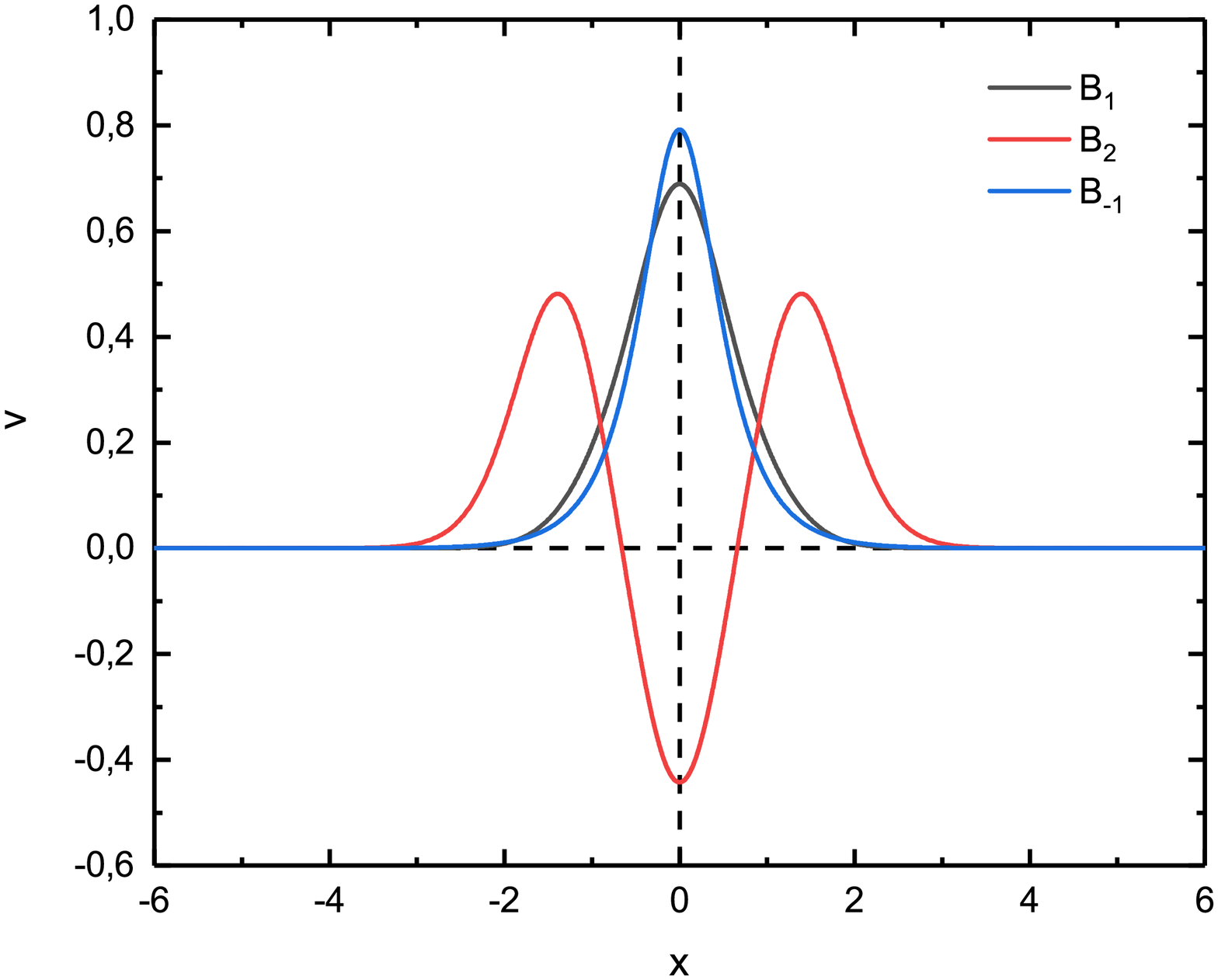}
        \includegraphics[width=.435\textwidth, trim = 40 20 90 20, clip = true]{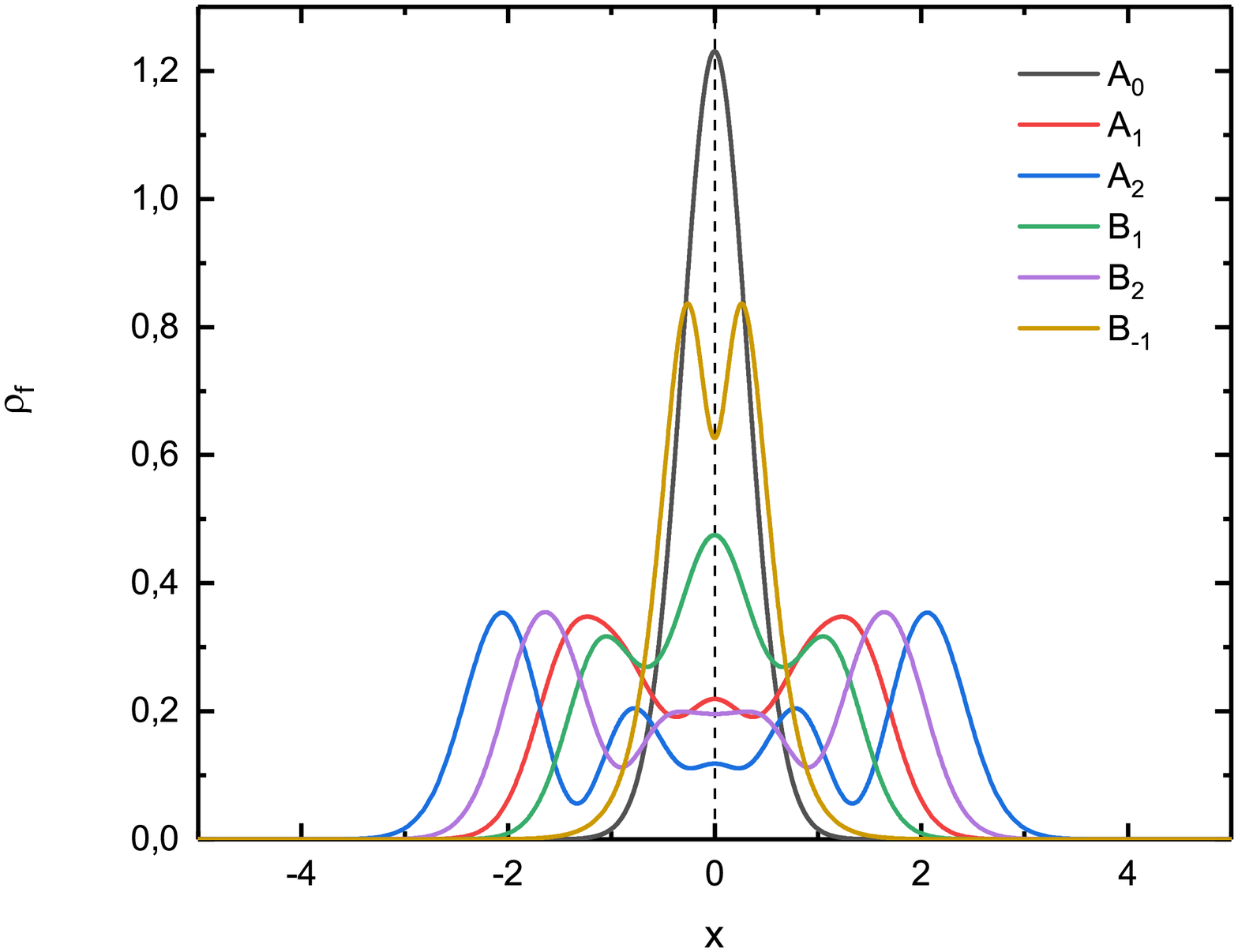}
    \end{center}
    \caption{\small
Components of the localized fermionic modes  of the types
$A_k$ (upper row), $B_k$ (middle row) and the fermion density distribution of these modes (bottom plot) are plotted
as functions of the coordinate $x$ for $m=0$ and $g=5$.
Backreaction of the fermions on the kink is taken into account.
}
\lbfig{Fig2}
\end{figure}

The situation is very different in the full coupled system \re{eqs} with backreaction,
the profile of the kink deforms as
a fermion occupies an energy level. Further, the energy levels of the bounded fermions move accordingly and the
symmetry between the localized fermion states with positive and negative eigenvalues $\epsilon$ is violated, see
Fig.~\ref{Fig1}, right plot.

Considering the spectral flow of the localized fermions, we observe that in the range of values of
the coupling $0<g<1$ there is only one localized zero mode, as seen in Fig.~\ref{Fig1}.
As the coupling $g$ grows,
we obtain an infinite tower of new solutions of the model \re{lag} which correspond to the states of
deformed $\phi^4$ kink with different types and filling factors of localized fermions with non-zero eigenvalues $\epsilon$.
The deformation of the coupled configuration drives the non-zero eigenvalues $\epsilon$ of all modes
towards negative energy continuum, see Fig.~\ref{Fig1}, right plot.

In Fig.~\ref{Fig2} we display the components of a few modes of both types, localized on the kink with backreaction,
and the corresponding
distributions of the fermionic density $\rho_f(x)$.

\begin{figure}[t!]
    \begin{center}
        \includegraphics[width=.435\textwidth, trim = 40 20 90 20, clip = true]{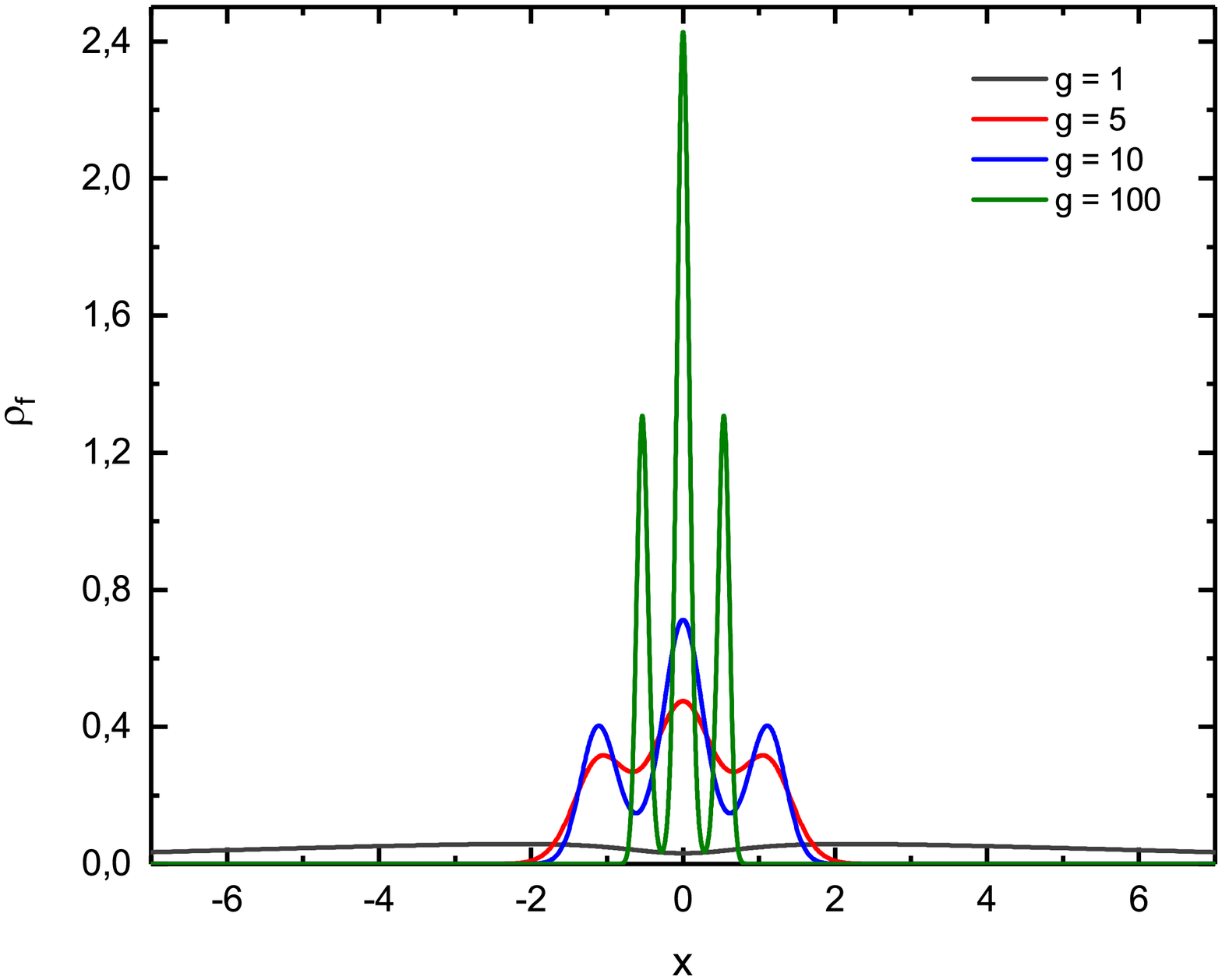}
        \includegraphics[width=.435\textwidth, trim = 40 20 90 20, clip = true]{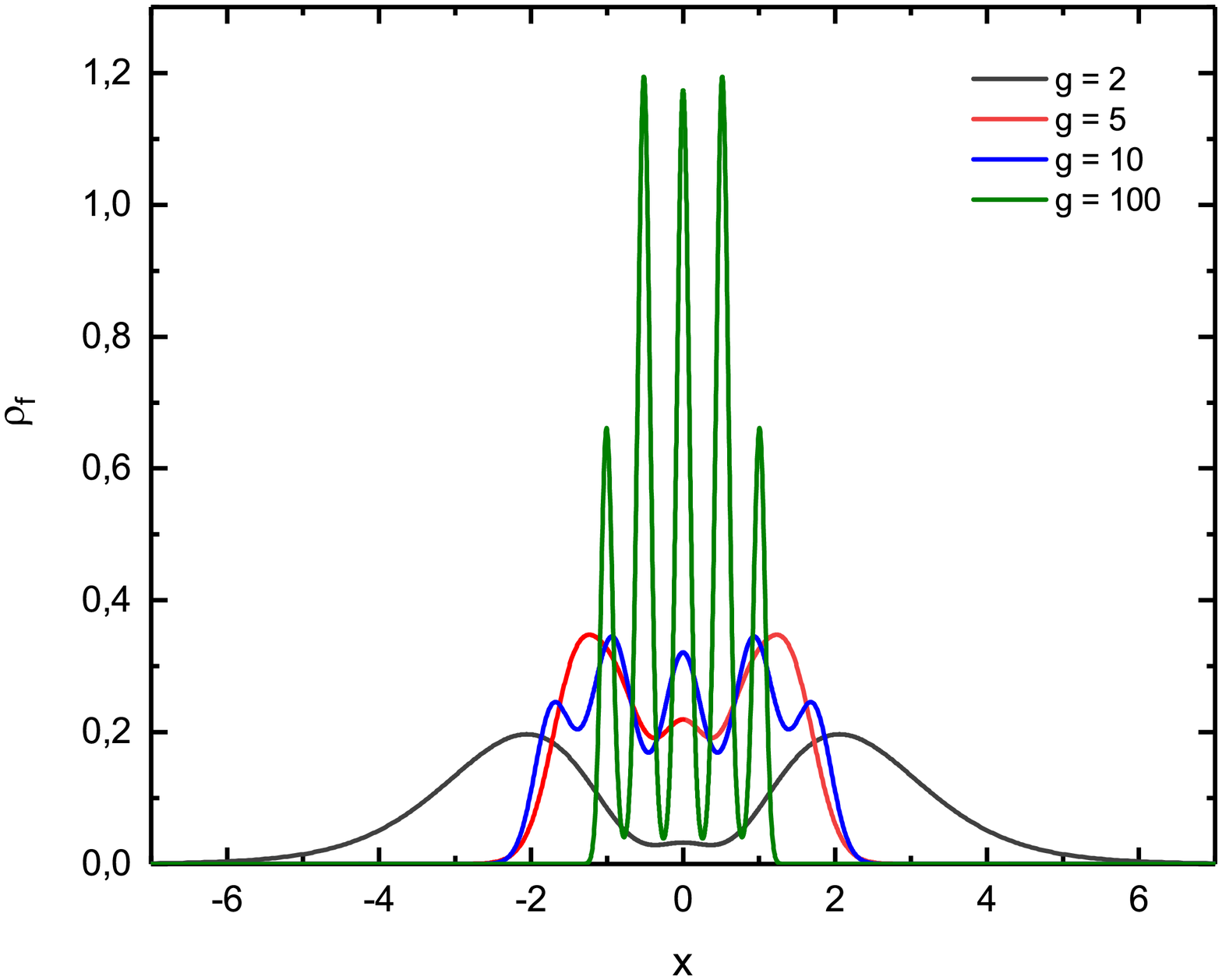}
        \includegraphics[width=.435\textwidth, trim = 40 20 90 20, clip = true]{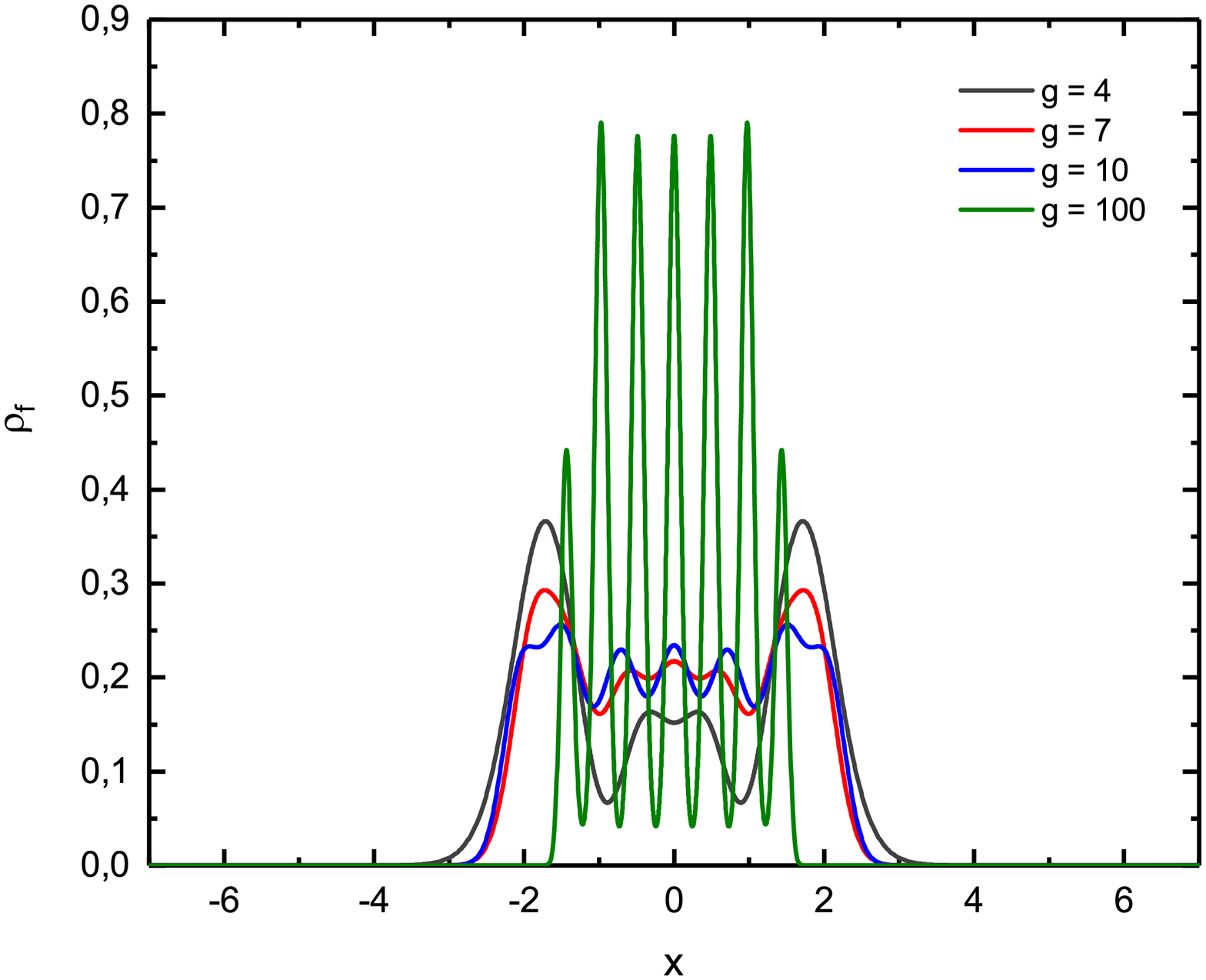}
        \includegraphics[width=.435\textwidth, trim = 40 20 90 20, clip = true]{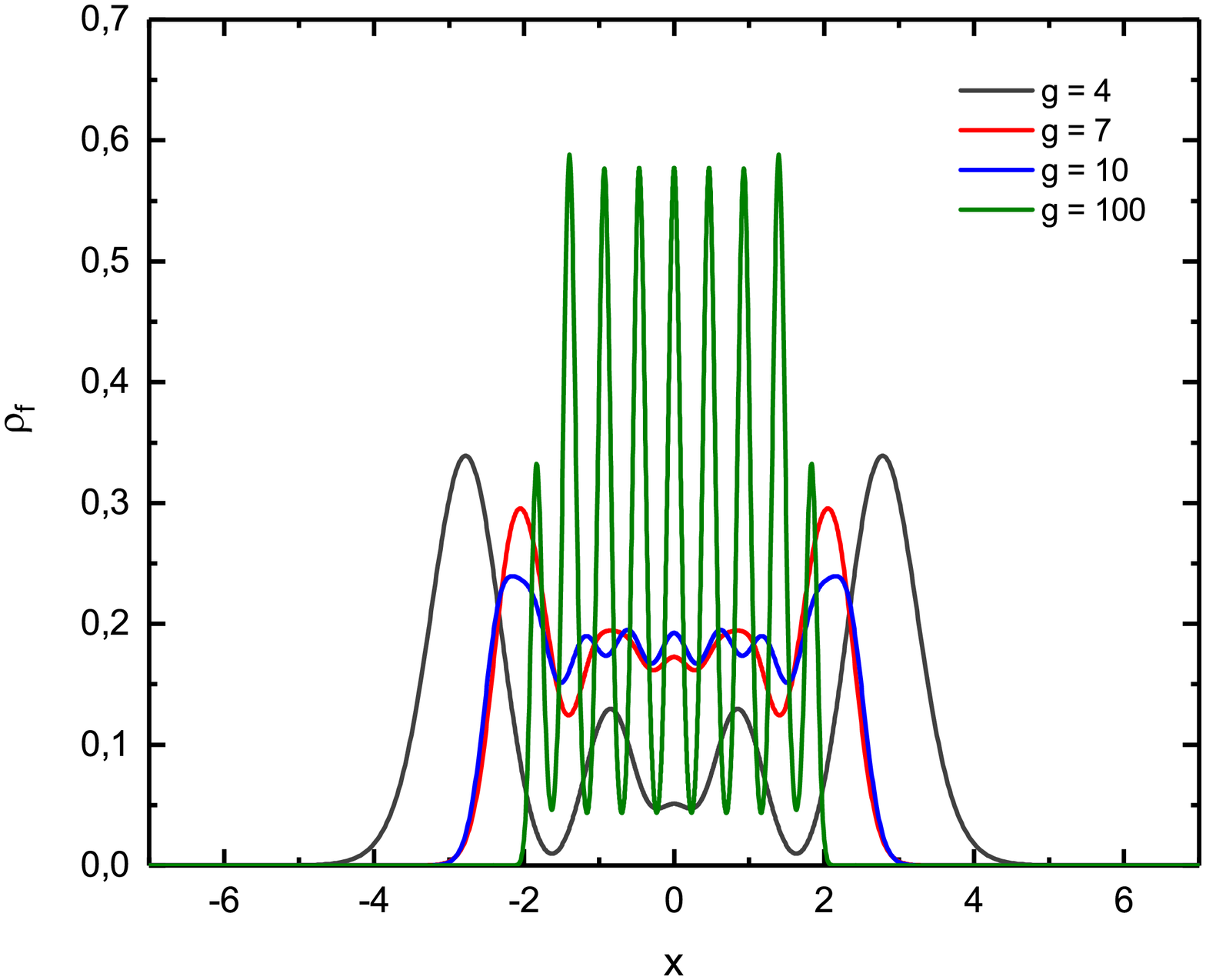}
    \end{center}
    \caption{\small
The fermion density distributions of the localized modes
$B_1$ (upper left plot), $A_1$ (upper right plot),
$B_2$ (bottom left plot) and $A_2$ (bottom right plot)  are plotted
as functions of the coordinate $x$ for $m=0$ and several values of the Yukawa coupling $g$.
Backreaction of the fermions on the kink is taken into account.
}
\lbfig{Fig3}
\end{figure}

In Fig.~\ref{Fig3} we plot the fermionic density distributions $\rho_f$
of the first four localizing modes $A_1,A_2,B_1,B_2$ with non-zero energy
for several values of coupling constant $g$.
At small values of the coupling constant $g$ there is only one localizing mode $A_0$
which should exists according to
the index theorem. As the coupling $g$ increases,
more modes of both types become localized by the kink.

Further increase of the coupling $g$ yields stronger bounding of the modes, also
larger number of localized modes are extracted from the positive and negative continuum.

The effect of backreaction of the fermions coupled to the kink is illustrated in Fig.~\ref{Fig4}.
As expected, the massless zero mode does not distort the kink for any values of the Yukawa coupling.
However, the scalar field is strongly affected by other bounded modes
with non-zero energy. For example, it is seen in Fig.~\ref{Fig4}, left upper plot, that
the coupling of the kink to the mode $B_1$ leads to
distortion of the profile of the soliton, which closely resembles the deformation of the kink
due to excitation of its normalizable discrete vibrational mode, see e.g. \cite{Shnir2018,Panos,Rajaraman:1975ez}.
Clearly, by analogy with excitations of this internal mode of the kink \cite{Manton:1996ex},
dynamical coupling to the fermions may lead to production of the kink-antikink pairs.

Coupling of the kink to the modes with some number of nodes is reflected in
visible spatial oscillations of the static scalar field at the center of the kink,
where fermion modes are located, see  Fig.~\ref{Fig4}. In some sense this configuration
can be thought of as a chain of kink-antikink pairs tightly bounded by the localized fermions.
Clearly, the coupling to higher fermionic modes yield much stronger
deformations of the kink.

\begin{figure}[t]
    \begin{center}
        \includegraphics[width=.435\textwidth, trim = 40 20 90 20, clip = true]{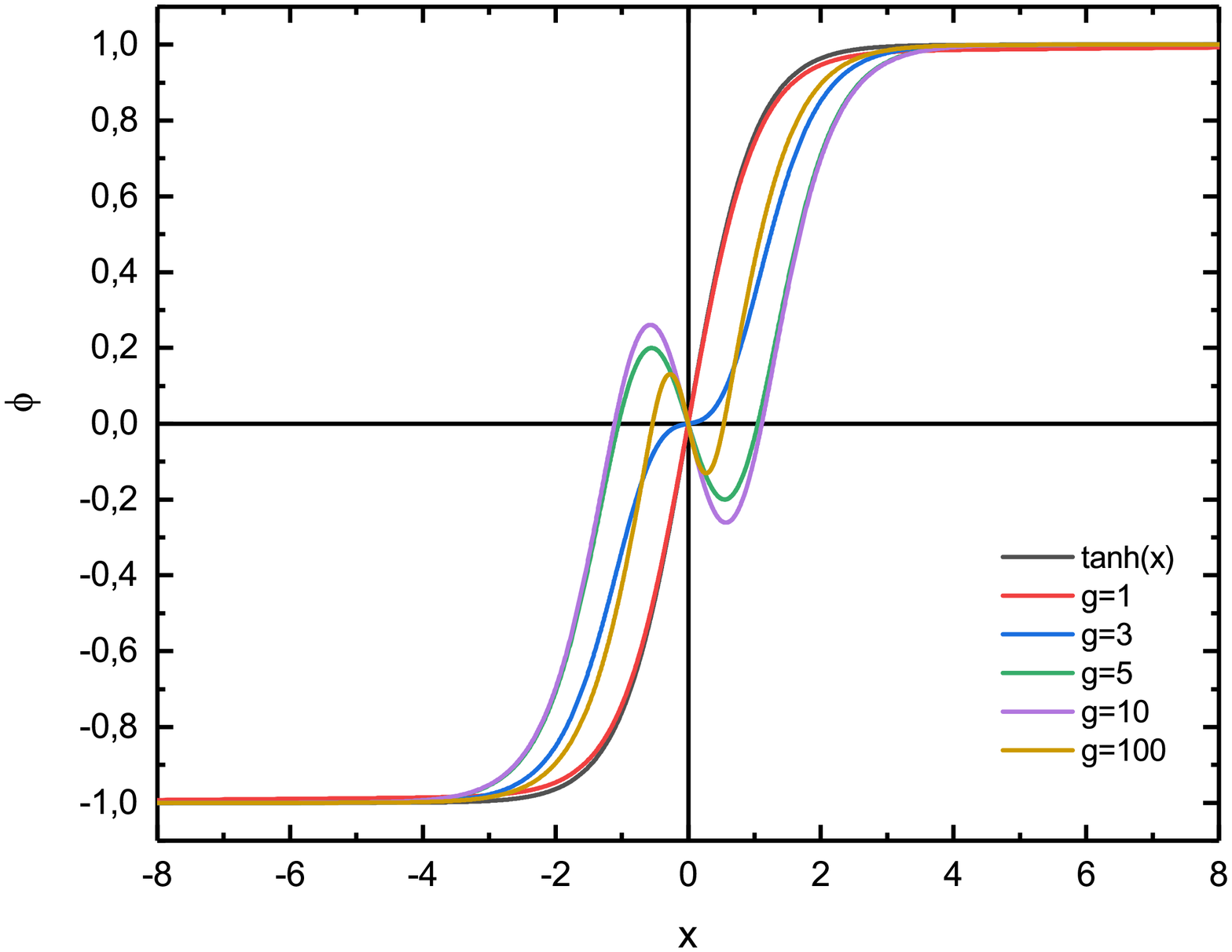}
        \includegraphics[width=.435\textwidth, trim = 40 20 90 20, clip = true]{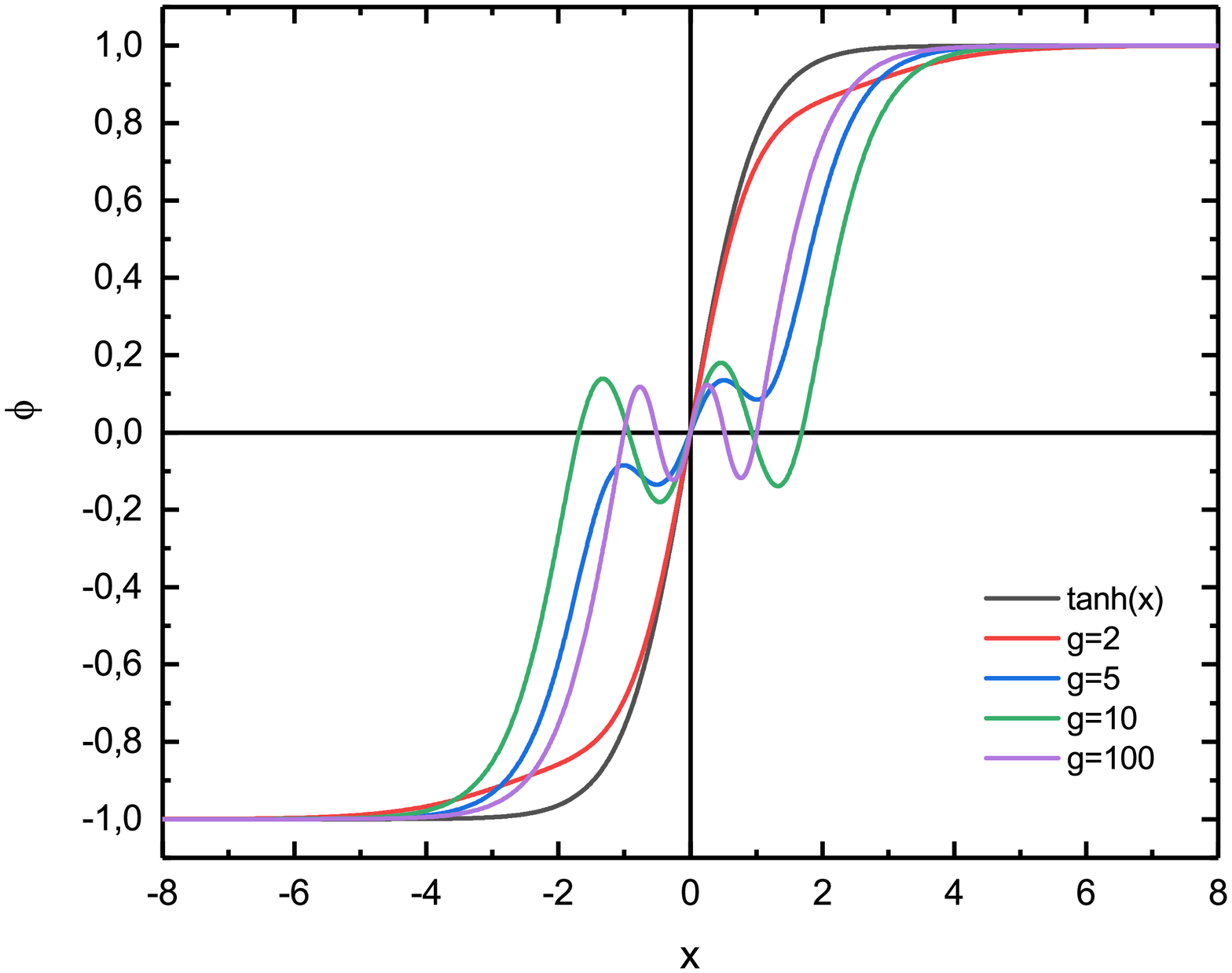}
        \includegraphics[width=.435\textwidth, trim = 40 20 90 20, clip = true]{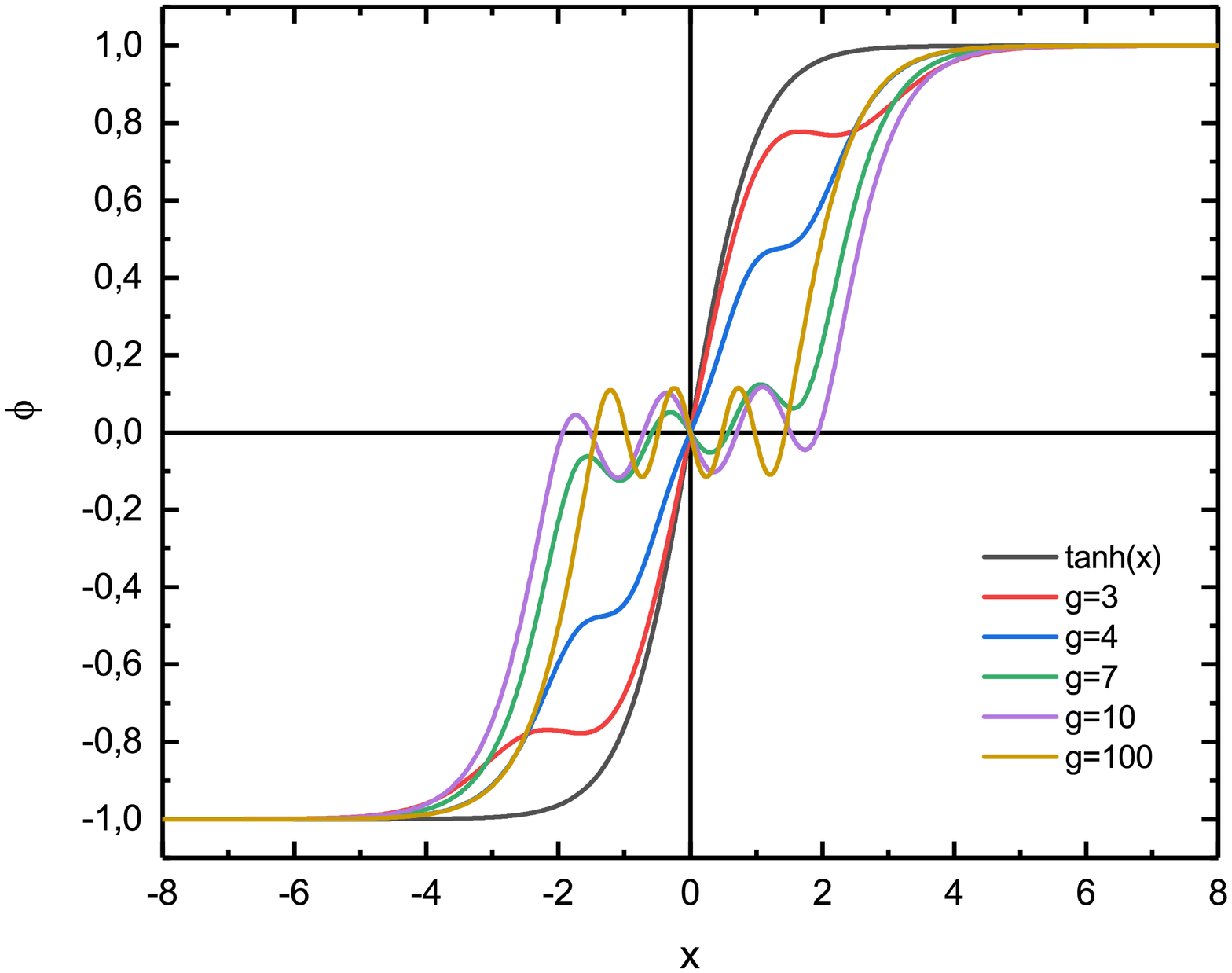}
        \includegraphics[width=.435\textwidth, trim = 40 20 90 20, clip = true]{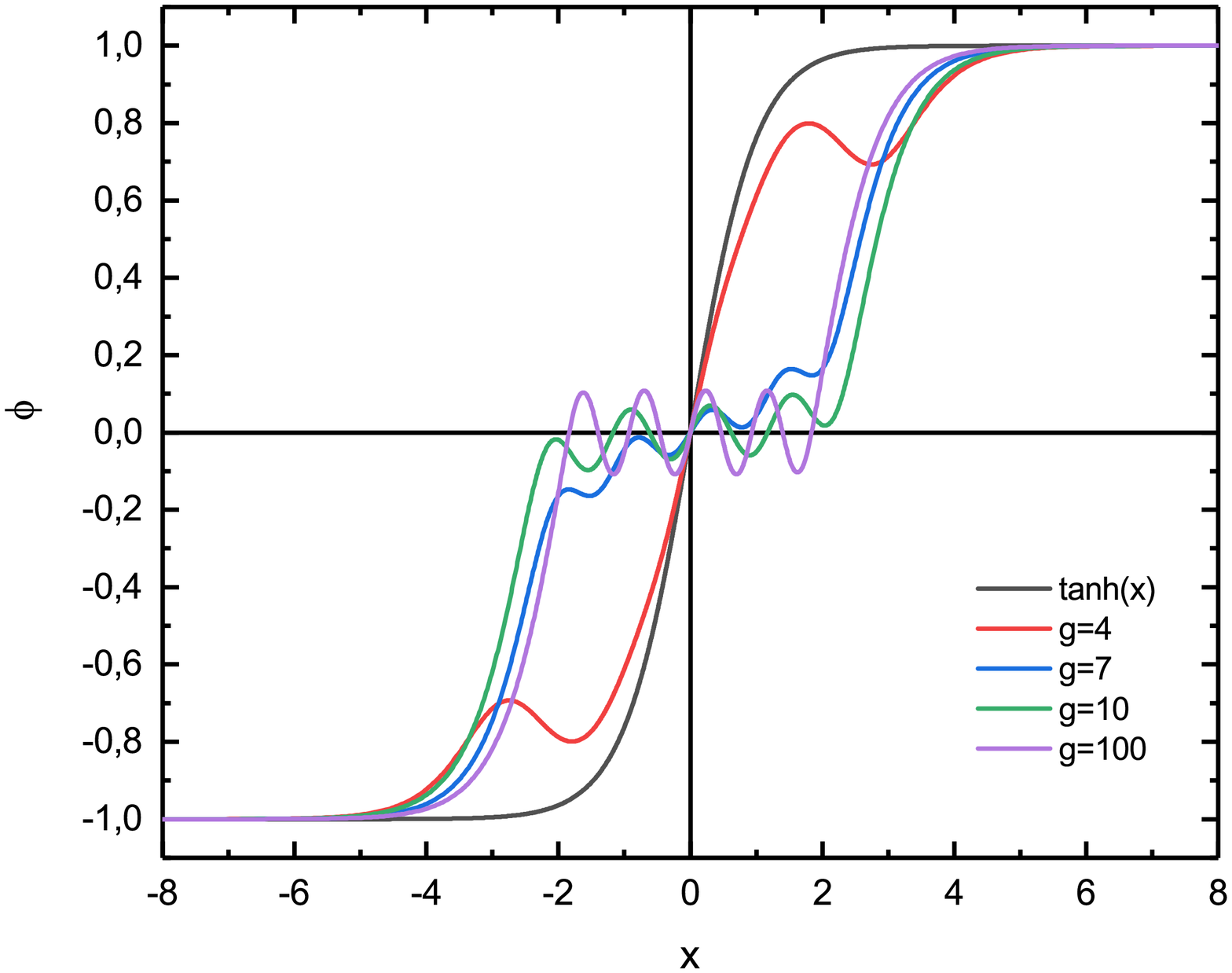}
    \end{center}
    \caption{\small
The profiles of the scalar field of the kink, coupled to the localized fermionic modes
$B_1$ (upper left plot), $A_1$ (upper right plot),
$B_2$ (bottom left plot) and $A_2$ (bottom right plot) for $m=0$ and
several values of the Yukawa coupling $g$.
}
\lbfig{Fig4}
\end{figure}

Next, we consider dependence of solutions on the value of the fermion bare mass $m$.
In that case the energy of the localized fermionic states is restricted as $| \varepsilon | < |g-m| $.
Note that, even for the fermions localized on the kink without the backreaction, the bare mass term violates the
reflection symmetry of the equations \re{eq3},  the term $m+g \tanh x$  does not have a definite parity.
Indeed, our numerical simulations confirm that for non-zero values of the bare mass $m$,
both the scalar field of the kink and fermionic densities of localized modes are asymmetric, see Figs.~\ref{Fig5},\ref{Fig6}
were we display numerical solution of the system \re{eq2} at $m=1$ for a fixed value of the coupling constant $g$.
Further increase of the fermion mass stronger deforms the configuration, the size of the deformed area increases.

The massive mode $A_0$ also becomes more localized, however, corresponding eigenvalue
remains zero and there is no zero-crossing mode in the spectral flow of the massive fermions
coupled to the kink in 1+1 dimensional system.
The spectrum of the massive fermions in not very different from the massless case above, we can see it
comparing figures ~\ref{Fig1} and Fig.~\ref{Fig7}, left plot.
For a fixed value of the Yukawa coupling the massive modes delocalize at some critical values of the fermion mass $m$,
see Fig.~\ref{Fig7}, right plot.

\begin{figure}[t!]
    \begin{center}
        \includegraphics[width=.435\textwidth, trim = 40 20 90 20, clip = true]{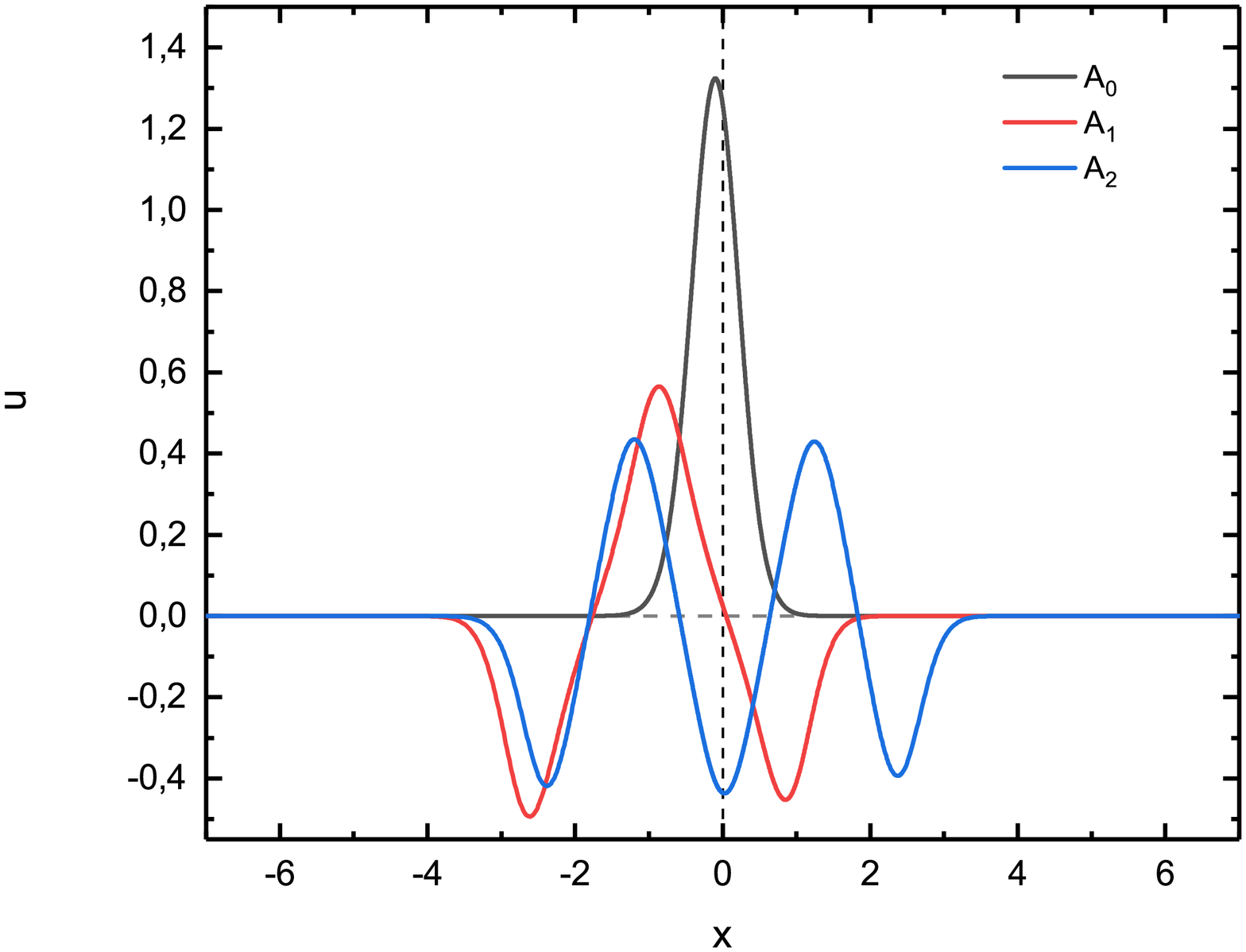}
        \includegraphics[width=.435\textwidth, trim = 40 20 90 20, clip = true]{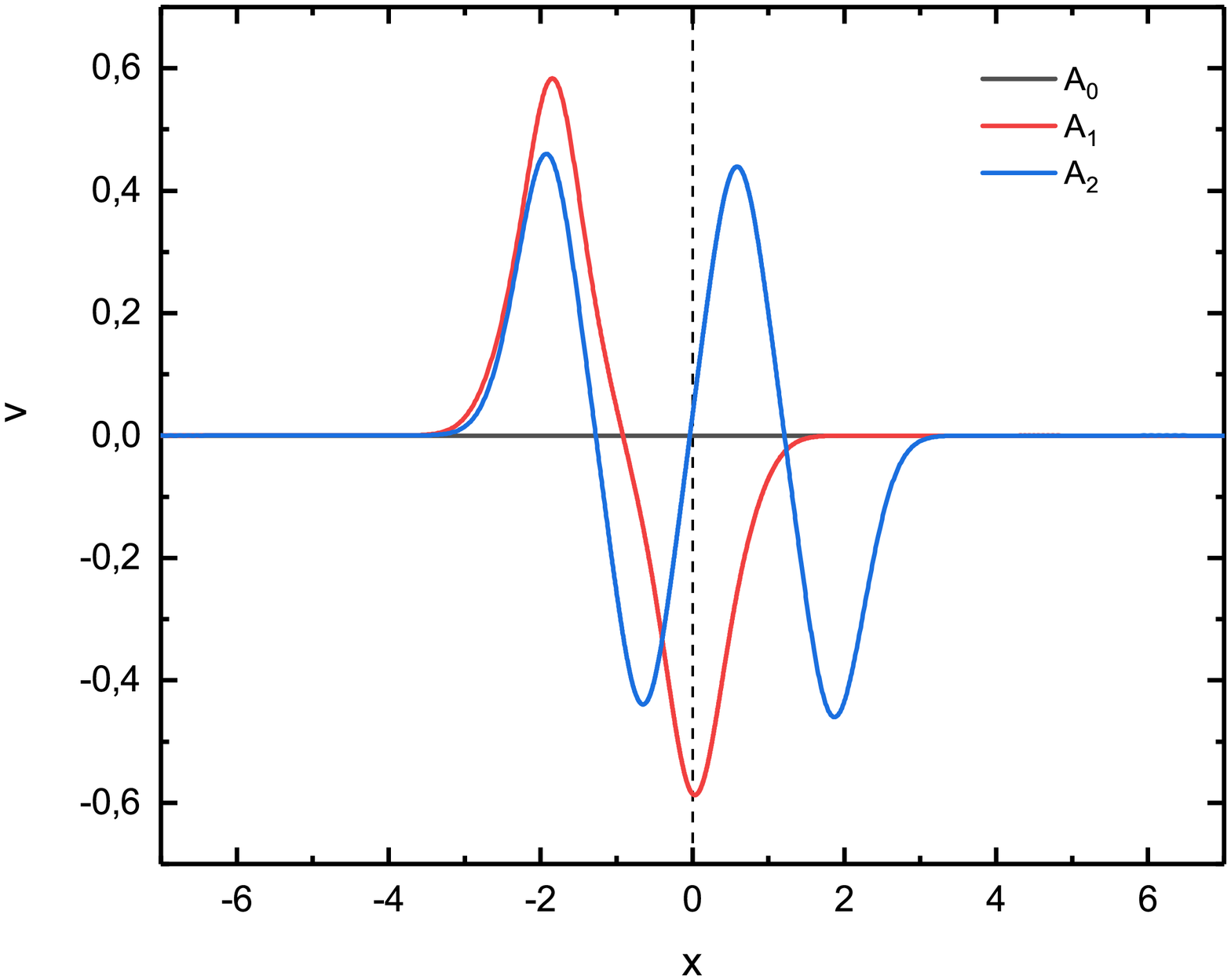}
        \includegraphics[width=.435\textwidth, trim = 40 20 90 20, clip = true]{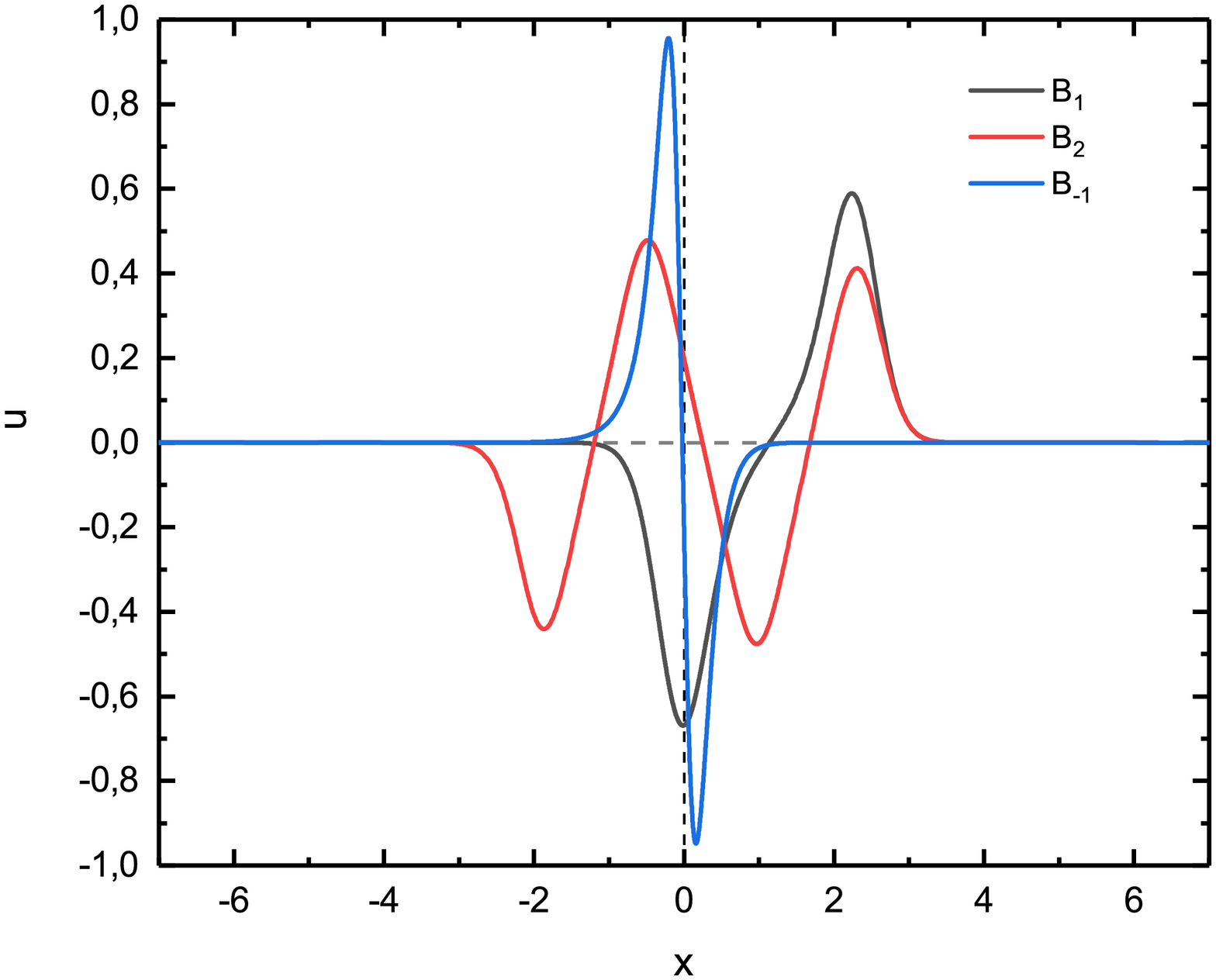}
        \includegraphics[width=.435\textwidth, trim = 40 20 90 20, clip = true]{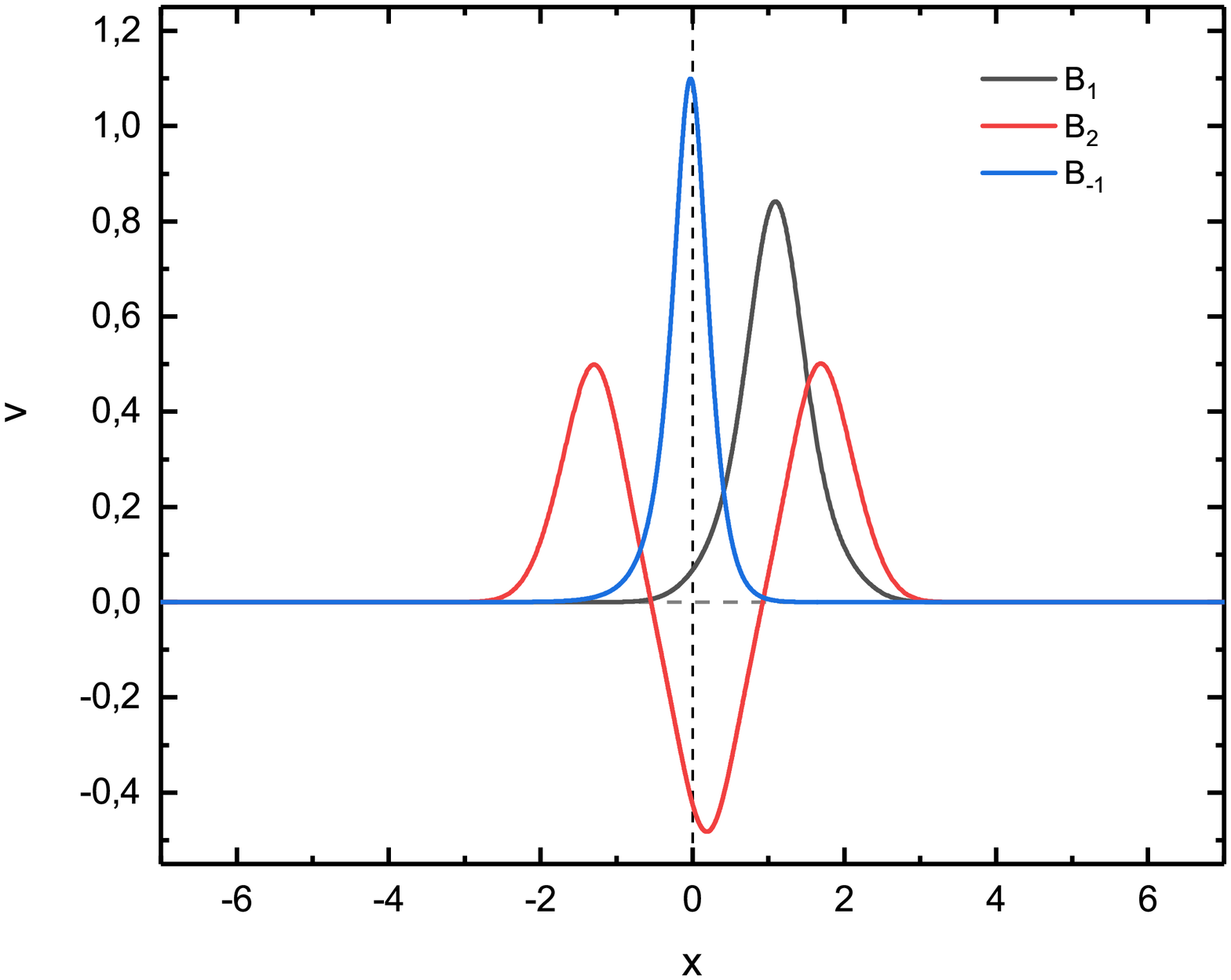}
        \includegraphics[width=.435\textwidth, trim = 40 20 90 20, clip = true]{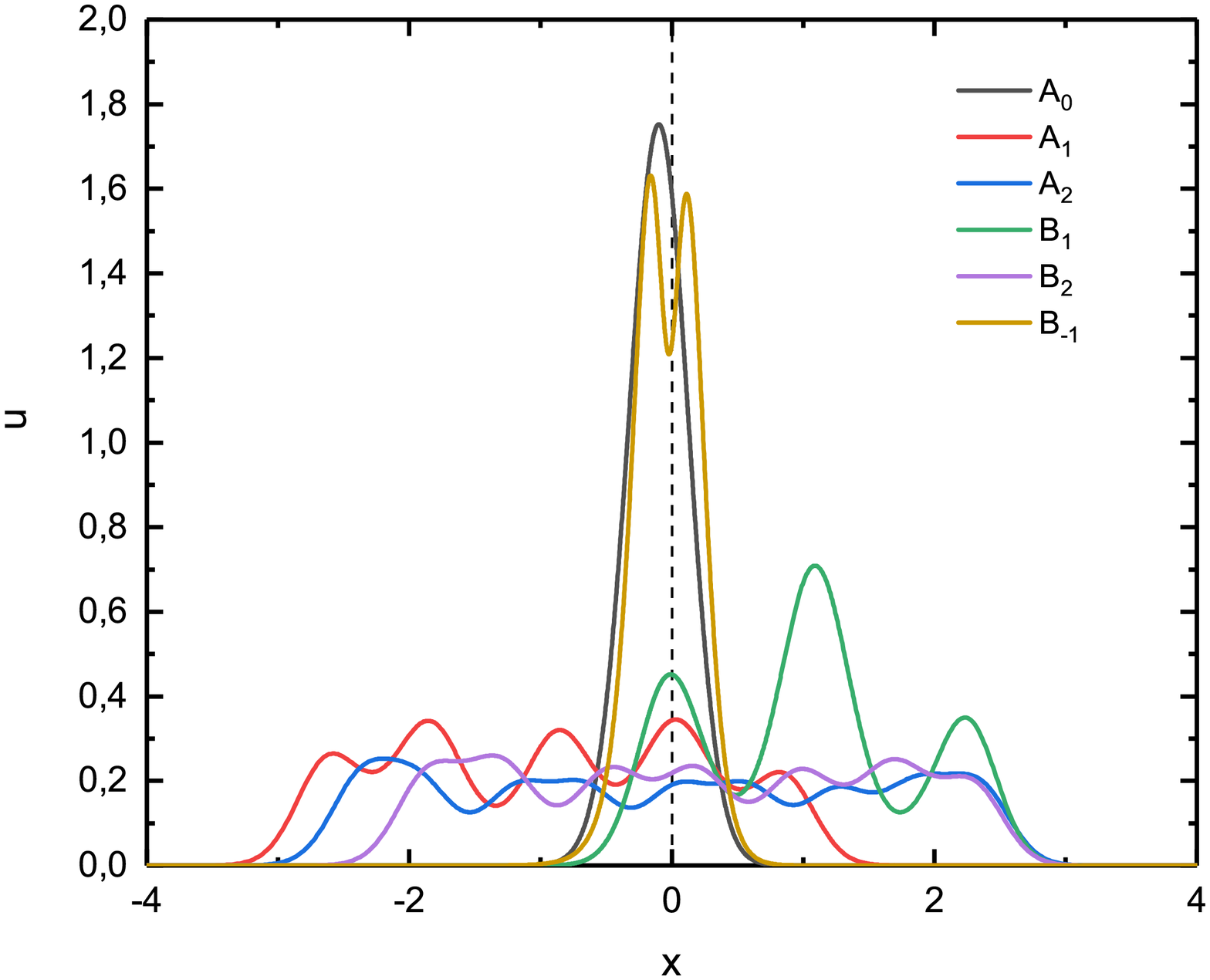}
    \end{center}
    \caption{\small
Components of the localized fermionic modes  of the types
$A_k$ (upper row) and $B_k$ (middle row) and the fermion density distribution of
these modes (bottom plot) are plotted
as functions of the coordinate $x$ for $m=1$ and $g=10$.
Backreaction of the fermions on the kink is taken into account.
}
\lbfig{Fig5}
\end{figure}

\begin{figure}[t]
    \begin{center}
        \includegraphics[width=.435\textwidth, trim = 40 20 90 20, clip = true]{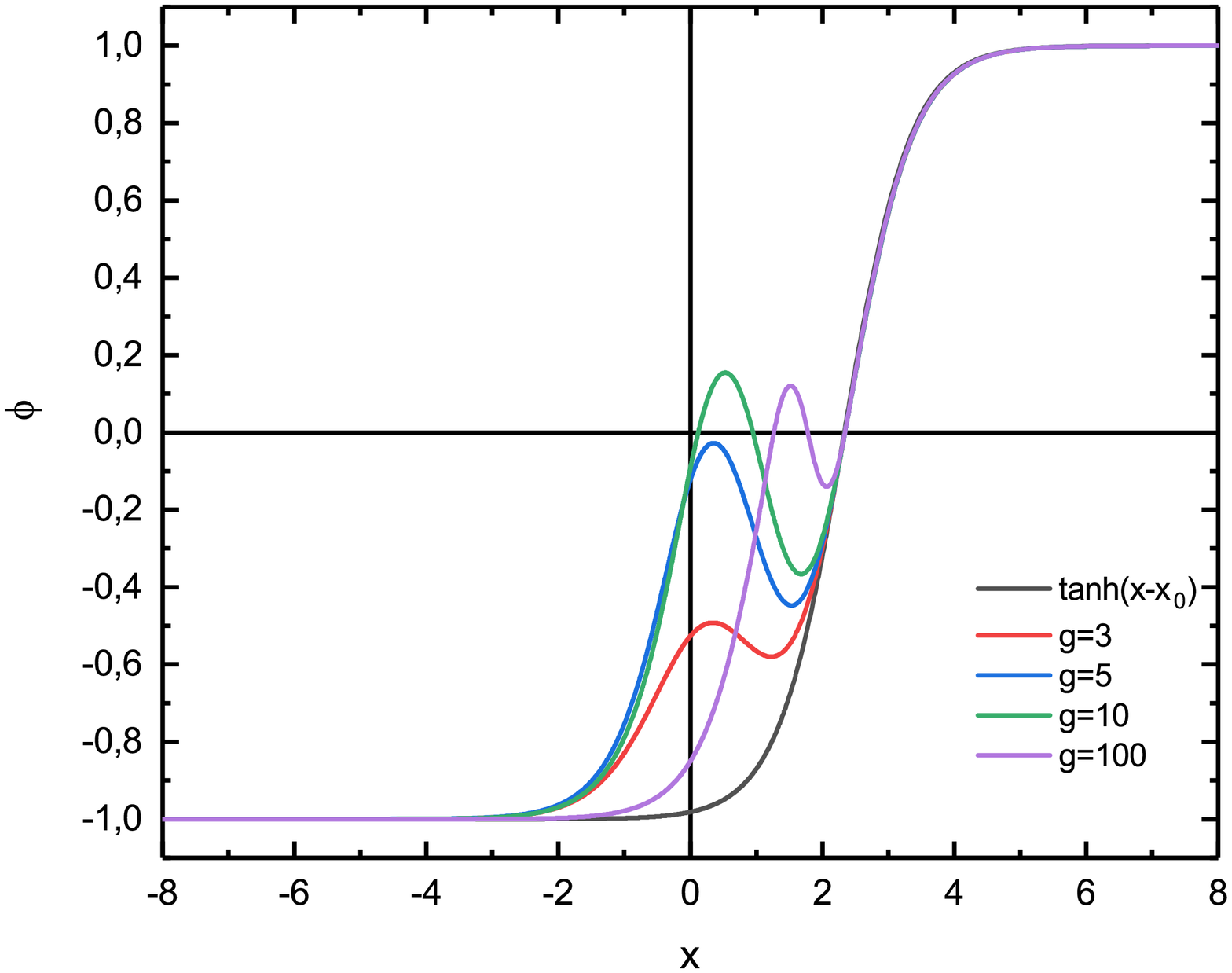}
        \includegraphics[width=.435\textwidth, trim = 40 20 90 20, clip = true]{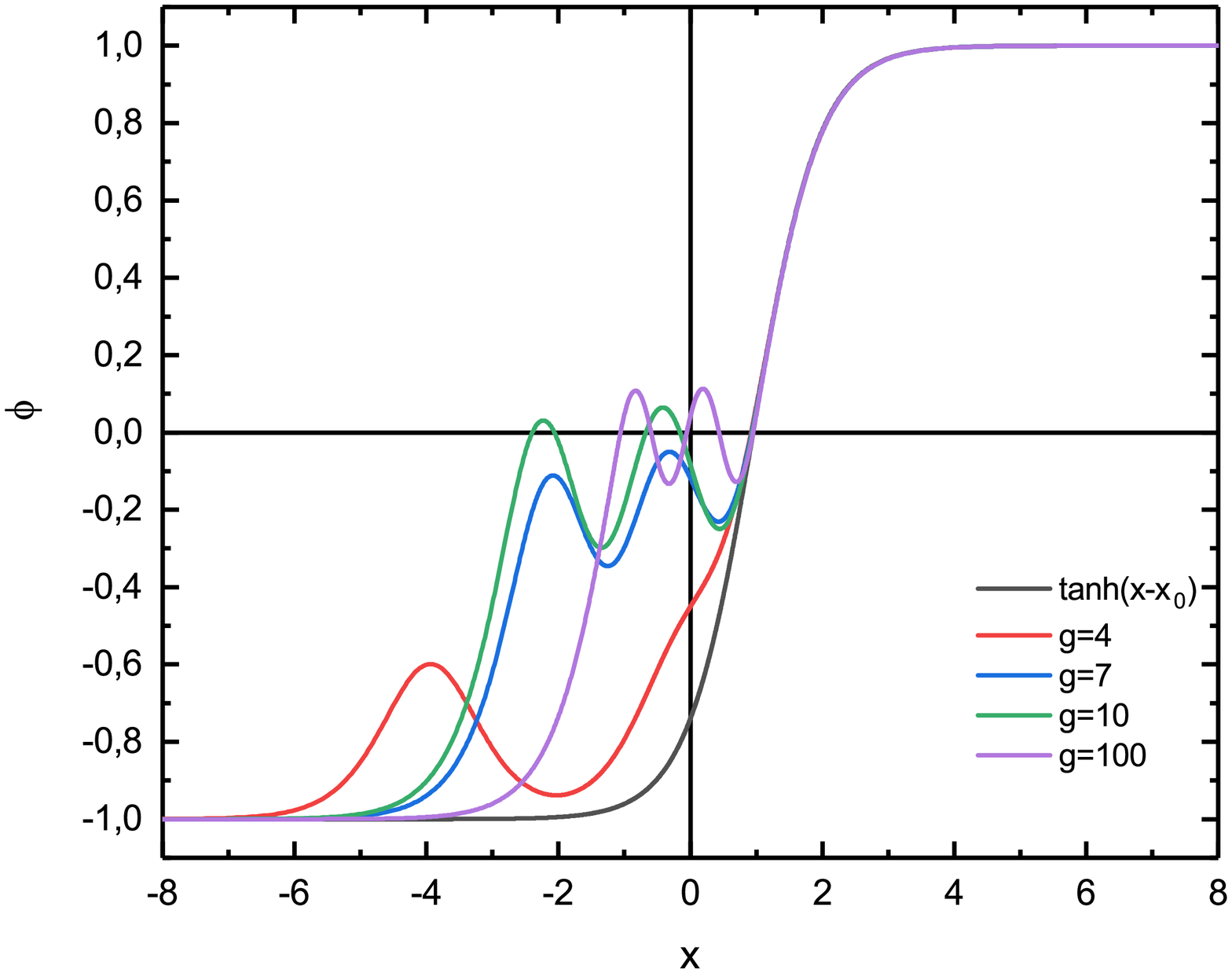}
        \includegraphics[width=.435\textwidth, trim = 40 20 90 20, clip = true]{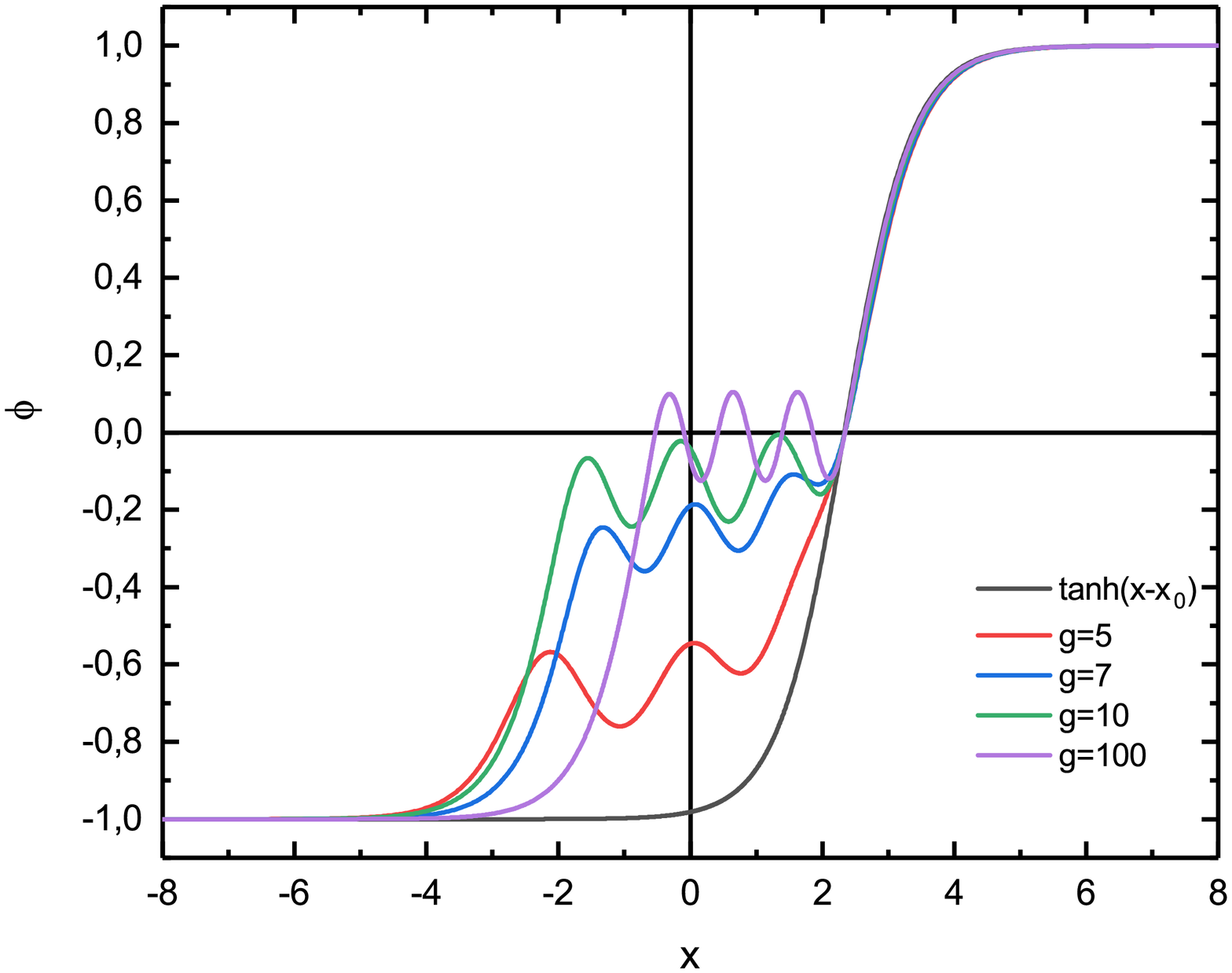}
        \includegraphics[width=.435\textwidth, trim = 40 20 90 20, clip = true]{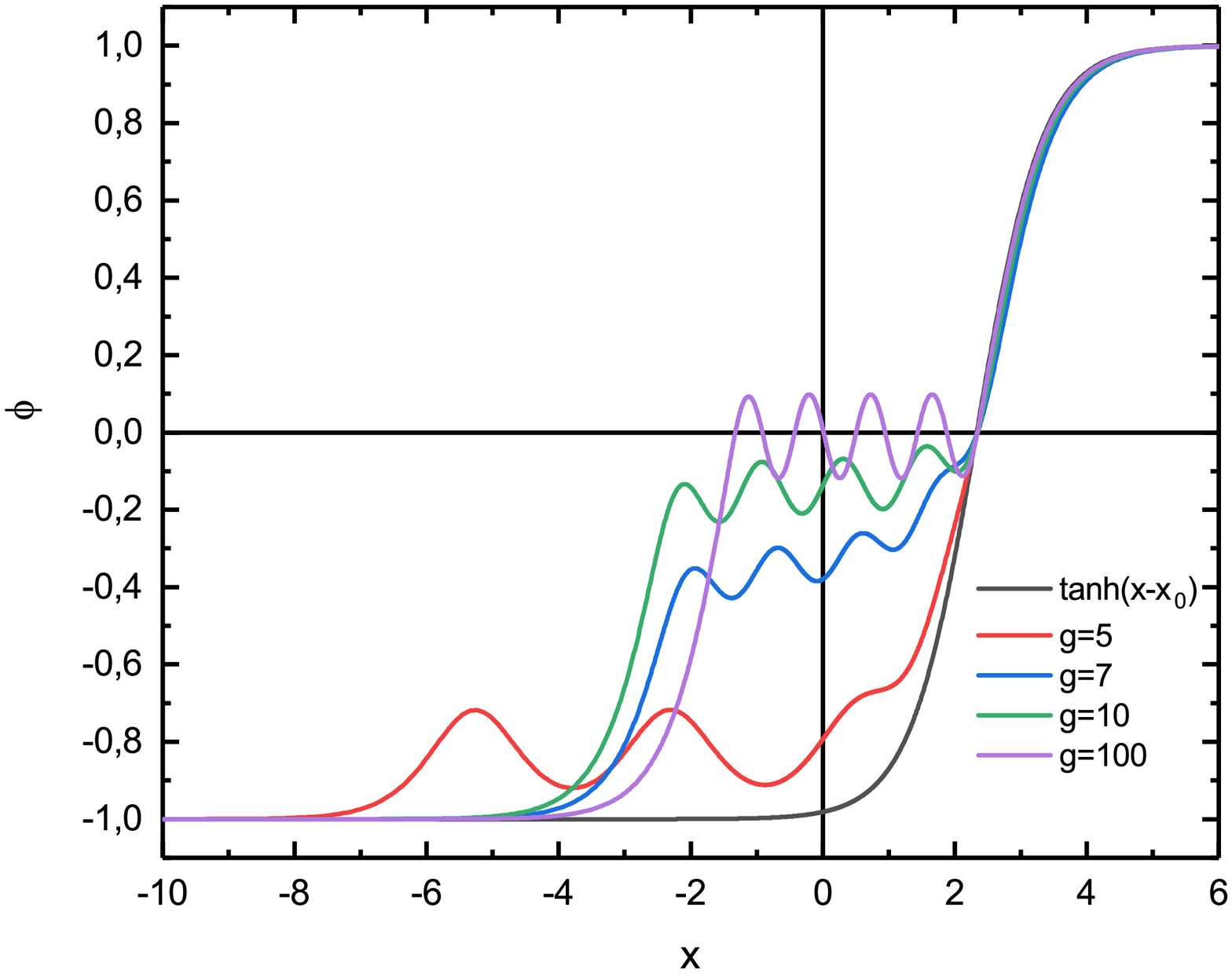}
    \end{center}
    \caption{\small
The profiles of the scalar field of the kink, coupled to the localized fermionic modes
$B_1$ (upper left plot), $A_1$ (upper right plot),
$B_2$ (bottom left plot) and $A_2$ (bottom right plot) for $m=1$ and
several values of the Yukawa coupling $g$.
}
\lbfig{Fig6}
\end{figure}

\begin{figure}[t]
    \begin{center}
        \includegraphics[width=.435\textwidth, trim = 40 20 90 20, clip = true]{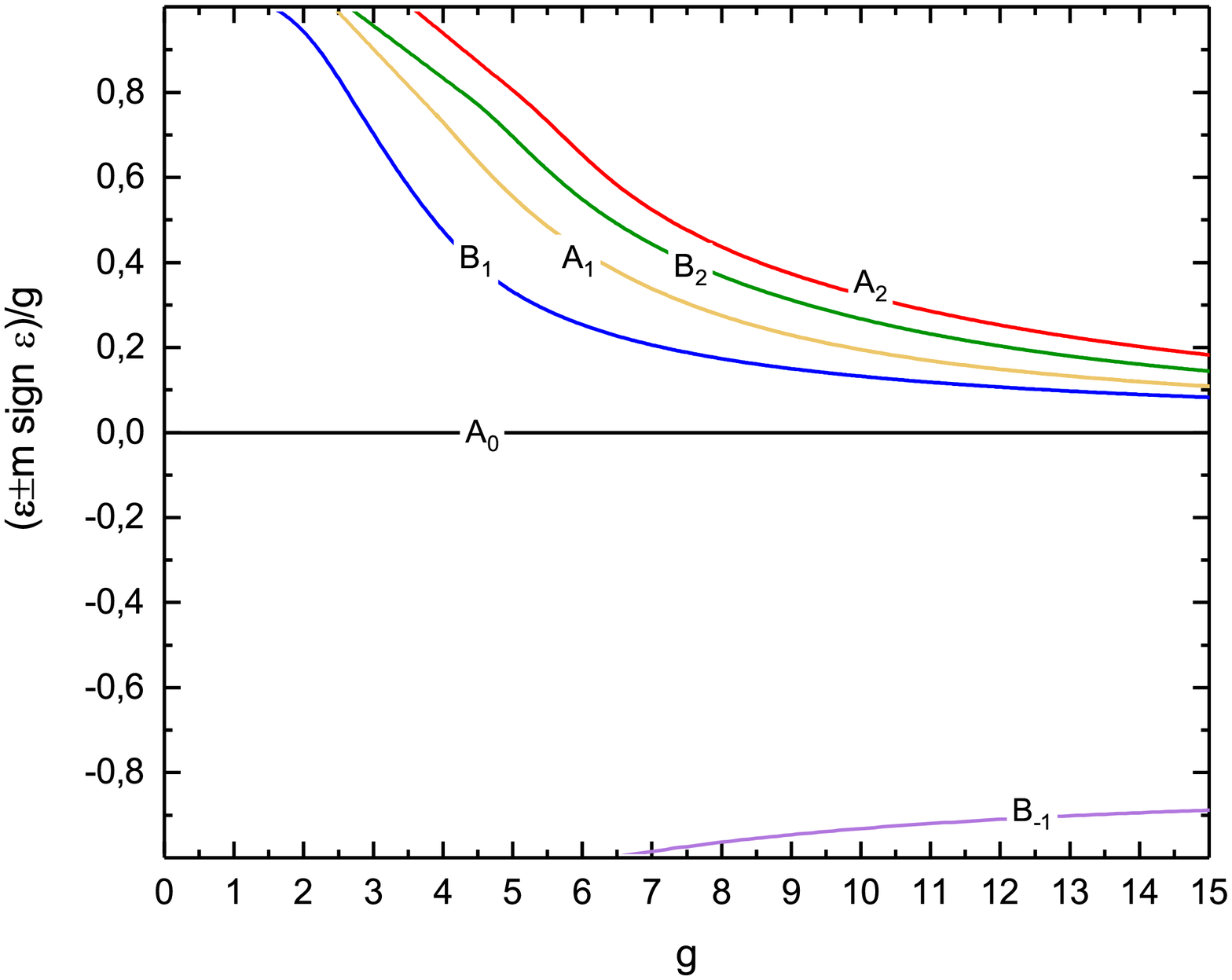}
        \includegraphics[width=.435\textwidth, trim = 40 20 90 20, clip = true]{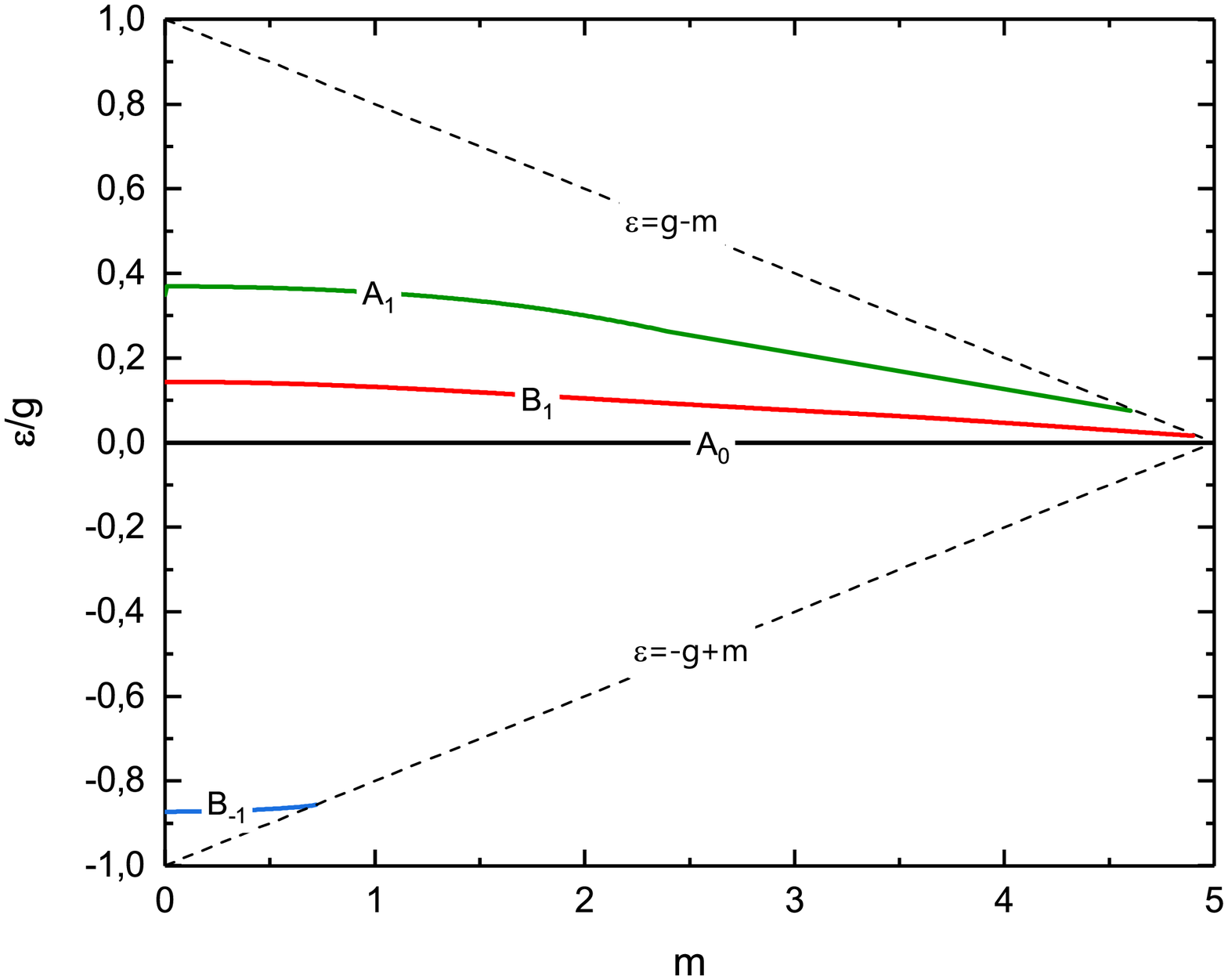}
    \end{center}
    \caption{\small
Normalized energy  $\frac{\epsilon + m ~{\rm sign}~\epsilon }{g}$ of the localized fermionic states
as a function of the Yukawa coupling $g$
for several fermion modes at $m=1$ (left plot) and the
normalized spectral flow  ${\epsilon }/{g}$ of the massive localized fermions as a function of the
bare mass $m$ (right plot).
}
\lbfig{Fig7}
\end{figure}

\section{Summary and conclusions}
In this paper we present a novel self-consistent
approach of analyzing fermionic states localized on the kink in the $\phi^4$ theory taking into account the
deformations of the soliton due to the presence of bounded fermions.
The full system of coupled field equations for the real scalar field and the
Dirac fermions coupled to the kink via the usual Yukawa coupling is supplemented by the normalization condition
for the localized fermions. Imposing appropriate boundary conditions both for scalar and spinor fields we
constructed numerical solutions of the resulting system of integral-differential
equations and found the corresponding energy eigenvalues. Apart the usual zero mode, which does not affect the
kink for any values of the coupling and the bare mass $m$, there is a tower of localized states with
non-zero eigenvalues, they are linked to the positive and negative continuum.

We have shown that the backreaction of the fermions
strongly affects the spectrum, it breaks the symmetry between the localized modes with positive and negative
eigenvalues. Further, the refection symmetry of the usual Dirac equation on static kink
background is violated for the fermions with non-zero bare mass, these spinors do not possess definite parity.

Localization of the fermions on the kink yields spatial oscillations of the
static scalar field at the center of the kink, where fermion modes are located.
This configuration  can be thought of as a
chain of kink-antikink pairs tightly bounded by the localized fermions, the number of the constituents on the
chain increases for higher fermionic modes.

The work here should be taken further by considering fermionic states localized  on various
topological solitons with backreaction.  In particular, it would be interesting to study the effects of the
fermionic modes localized on non-BPS monopoles, and on the Abelian vortices and the corresponding superconducting strings.
Another interesting question, which we hope to be addressing in the near future, is to
investigate the exchange interaction between the solitons mediated by the fermions, a first step
in this direction has been made in \cite{Perapechka:2019dvc}.

\section*{Acknowledgements}
Ya.S. would like to thank Steffen Krusch, Tomasz Roma\'{n}czukiewicz and Andrzej Wereszczy\'{n}ski
for enlightening discussions. He gratefully acknowledges partial support of the Ministry of Science and
Higher Education of Russian Federation, project No~3.1386.2017.

\end{document}